\documentclass{article}

\usepackage{arxiv}

\usepackage[utf8]{inputenc} 
\usepackage[T1]{fontenc}    
\usepackage{hyperref}       
\usepackage{graphicx}       
\usepackage{xcolor}
\usepackage{natbib} 
\usepackage{amsfonts}       
\usepackage{amsmath}
\usepackage{amsthm}
\usepackage{bbm}
\usepackage{booktabs}
\usepackage{algorithm}
\usepackage{algpseudocode}
\usepackage{adjustbox}
\PassOptionsToPackage{export}{adjustbox}
\usepackage{multirow} 
\usepackage{array}
\usepackage{eurosym}
 
\usepackage[shortlabels]{enumitem}

\newtheorem{theorem}{Theorem}

\def\bSig\mathbf{\Sigma}

\usepackage{comment}
\usepackage{standalone}

\title{An interpretable single-index mixed-effects model for Non-Gaussian national survey data}

\newif\ifuniqueAffiliation
\uniqueAffiliationtrue

\ifuniqueAffiliation 
\author{ \href{https://orcid.org/0000-0003-3265-6330}{\includegraphics[scale=0.06]{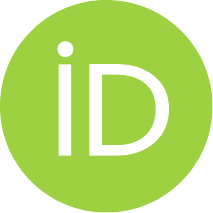}\hspace{1mm}Qingyang Liu} \\
	Department of Statistics\\
	University of Wisconsin-Madison\\
	Madison, WI 53706 \\
	\texttt{qliu432@wisc.edu} \\
	\And
	\href{https://orcid.org/0000-0002-5345-8635}{\includegraphics[scale=0.06]{orcid.pdf}\hspace{1mm}Debdeep Pati} \\
	Department of Statistics\\
	University of Wisconsin-Madison\\
	Madison, WI 53706 \\
	\texttt{dpati2@wisc.edu} \\
	\And
	\href{https://orcid.org/0000-0001-5421-1725}{\includegraphics[scale=0.06]{orcid.pdf}\hspace{1mm}Dipankar Bandyopadhyay} \\
	Department of Biostatistics\\
	Virginia Commonwealth University\\
	Richmond, VA 23219 \\
	\texttt{dbandyop@vcu.edu} \\
}

\begin{document}
\maketitle

\begin{abstract}
This manuscript presents an innovative statistical model to quantify periodontal disease in the context of complex medical data. A mixed-effects model incorporating skewed random effects and heavy-tailed residuals is introduced, ensuring robust handling of non-normal data distributions. The fixed effect is modeled as a combination of a slope parameter and a single index function, constrained to be monotonic increasing for meaningful interpretation. This approach captures different dimensions of periodontal disease progression by integrating Clinical Attachment Level (CAL) and Pocket Depth (PD) biomarkers within a unified analytical framework. A variable selection method based on the grouped horseshoe prior is employed, addressing the relatively high number of risk factors. Furthermore, survey weight information typically provided with large survey data is incorporated to ensure accurate inference. This comprehensive methodology significantly advances the statistical quantification of periodontal disease, offering a nuanced and precise assessment of risk factors and disease progression. The proposed methodology is implemented in the \textsf{R} package \href{https://cran.r-project.org/package=MSIMST}{\textsc{MSIMST}}.
\end{abstract}

\keywords{Single-Index Model, Robust, Heavy Tail, Skewness}

\section{Introduction} \label{sec:introduction}
Despite recent significant advances in preventive measures and strategies, such as water fluoridation and dental sealants, aimed at improving the oral health status of Americans, periodontal disease continues to remain a silent epidemic \citep{benjamin2010oral}. The complications associated with untreated periodontal disease include discomfort and pain, poor appearance, loss of self-esteem, difficulties in speaking, mastication, and swallowing, leading to impaired quality of life, eventual tooth loss, and potentially limited food choices, resulting in poor nutrition. As complex chronic diseases with distinct pathophysiologies, the manifestation and progression of periodontal disease are multifactorial. The ultimate goal of dental treatments is to prevent tooth loss and maintain the dentition in a state of comfort and function. However, the significant cost burden calls for the development of pragmatic tools for efficient risk evaluation of periodontal disease, which is also associated with several systemic non-communicable diseases, such as cardiovascular diseases, rheumatoid arthritis, and Type-2 diabetes, where the multi-comorbidity relation is perceived as bi-directional \citep{taylor2001bidirectional}.

To develop an adequate evaluation tool for periodontal disease studies, we must overcome five key challenges. First, most of the available complex statistical tools for periodontal disease studies often uncritically use Gaussian assumptions, leading to imprecise parameter estimates for highly right-skewed and heavy-tailed periodontal disease responses \citep{bandyopadhyay2010linearmixed}. Alternative transformations, such as the Box-Cox transformation, to achieve normality come with known practical difficulties, including determining the universally accepted class of transformation to (multivariate) normality and a lack of clinical interpretation of results at the original scales of the responses. Second, most existing models either impose stringent linearity assumptions between the covariates and the response variables or belong to “black box” models with poor interpretability. A flexible parametric or semi-parametric statistical model with high interpretability is more favorable than these models. Third, it is essential to include both pocket depth (PD) and clinical attachment level (CAL) as response variables since both are routinely measured in clinical practice and used to make treatment decisions. Fourth, leveraging large-scale surveys is necessary to comprehensively investigate the prevalence and determinants of periodontal disease across diverse demographic groups. An adequate evaluation tool must incorporate survey weight information often provided with large-scale surveys to ensure accurate inference and representation of the intrinsic complex sampling method. Last, the existence of a high number of risk factors in large-scale surveys necessitates the adoption of a variable selection method.

We overcome these five impediments one by one. First, we propose a Bayesian mixed-effect model that replaces the Gaussian assumption on the residual terms with the Student-$t$ distribution, which is suitable for potential heavy-tailed data. Additionally, we assume that the random effect term follows the Skew-$t$ (ST) distribution that belongs to the skew-normal/independent distribution family \citep{lachos2010likelihood,schumacher2021scale}. The ST distribution is flexible enough to model skewed, non-normal data. It includes parameters that capture skewness and heavy tails, and it formally encompasses the Gaussian, Student-$t$, and skew-normal (SN) distributions as special cases. Second, within the proposed model, we adopt a single index function that assumes the combined effect of the risk factors on a subject is captured by a scalar, the single index, which is a linear combination of the risk factors. The magnitude and direction of the coefficients determine the relative importance of the corresponding risk factor. Our proposed model generalizes the standard linear model by allowing the mean response to be a general \emph{non-linear} function of the single index and allowing the residuals to be non-Gaussian. For interpretability, the index function is restricted to be a monotonic increasing function of the single index, allowing the index to rank patients according to their risk of periodontal disease. Third, integrating PD and CAL into a comprehensive model offers a more holistic view and captures the multifaceted nature of periodontal diseases. To achieve this, we “stack” the fixed effect terms of PD and CAL and introduce a slope parameter to account for the association between PD and CAL. Fourth, large-scale surveys generally have complex sampling methods and have supplied survey weight information to represent the intrinsic complex sampling method. To incorporate survey weight information into the proposed model, we adopt the methodology from \citet{Gunawan2020}, which applies to complex Bayesian models like the one we propose and ensures accurate inference and representation of the intrinsic complex sampling method used in survey data. Last, we tackle the challenge of a high number of risk factors in large-scale surveys by adopting the grouped horseshoe prior within the proposed model for its satisfying empirical performance in variable selection \citep{carvalho2010horseshoe}. 

We summarize the main contributions of this paper as follows:
\begin{enumerate}[{(1)}]
	\item We introduce a single index mixed-effects model with skewed random effects and heavy-tailed residuals, designed explicitly for quantifying periodontal disease. We call this model the ST-GP model. The rationale behind the name ST-GP model will be elaborated in Section \ref{sec:methodology}. The ST-GP model incorporates a monotonic increasing single index function \emph{without} the linearity assumption. Notably, the ST-GP model integrates both PD and CAL as response variables, thereby removing the necessity of fitting separate models for PD and CAL.
	\item We adopt a Bayesian procedure from \citet{Gunawan2020} to incorporate survey weight information supplemented with large survey data. Failing to incorporate the intrinsic sampling mechanism in the survey data would lead to inconsistent estimation of covariate coefficients and erroneous inference results.
	\item We employ a grouped variable selection prior (the grouped horseshoe prior) to facilitate variable selection. The number of covariates is commonly large for large survey data, making a shrinkage prior necessary for separating the signal from the noise.
	\item We propose a tuning-free Gibbs sampler for the ST-GP model. Existing Bayesian single index models often use traditional samplers such as the Metropolis-Hasting algorithm or the reversible-jump Markov chain Monte Carlo algorithm \citep{antoniadis2004bayesian,wang2009bayesian,choi2011gaussian,gramacy2012gaussian}, which require careful tuning such as step sizes or proposal distributions, which can be both time-consuming and prone to error. Our tuning-free approach removes this burden, allowing users to focus on model development and interpretation rather than algorithmic intricacies.
\end{enumerate}

\subsection{Literature Review}

Various mixed-effects models are flexible and robust enough for non-Gaussian data. For instance, \citet{pinheiro2001efficient, Rosa2003} introduced linear mixed models with heavy-tailed and \emph{symmetric} random effects and residuals. Similarly, \cite{arellano2005skew, Ho2010, lachos2010likelihood} proposed linear mixed models featuring heavy-tailed residuals and \emph{asymmetric} random effects. \citet{bandyopadhyay2010linearmixed} suggested a Bayesian linear mixed model with skewed random effects and heavy-tailed residuals, specifically applied to stack biomarkers, PD and CAL, as the response variable. However, all these mixed-effect models belong to the linear models family and impose stringent linearity assumptions.  

In contrast, the single index model is capable of modeling non-linear relationships and is supported by extensive literature, including works by \citet{stoker1986consistent,ichimura1993semiparametric,carroll1997generalized,ruppert2002selecting,wang2009spline,Kuchibhotla2020}. For skewed data, several studies have extended the single index model within the quantile regression framework, including \citet{wu2010single,zhu2012semiparametric,Ma2016,gardes2018tail,xu2022extreme}. Additionally, \citet{pang2012estimation} proposed a single index model with random effects utilizing the generalized estimating equations method. However, these single index models have not addressed the issue of highly correlated response variables (PD and CAL) nor tackled the challenge related to the relatively high number of risk factors. To the best of our knowledge, the model we will introduce in this paper is the only one capable of overcoming all five key challenges simultaneously.

\subsection{Exploratory Data Analysis} \label{sec: Exploratory Data Analysis}

In this paper, we aim to provide nationwide estimates of periodontal disease in the United States, utilizing large, government-funded, nationally representative databases like the National Health and Nutrition Examination Survey (NHANES) spanning the years 2009 to 2014 \citep{NHANES}. NHANES offers extensive information on periodontal disease and comorbidities and stands out due to its comprehensive approach, including interviews and physical examinations. NHANES gathers data on various aspects, including the prevalence of chronic and infectious diseases and conditions, even those undiagnosed, along with risk factors such as obesity, elevated serum cholesterol levels, hypertension, dietary habits, nutritional status, and numerous other measures.

\begin{figure}
	\centering
	\includegraphics[width = \textwidth]{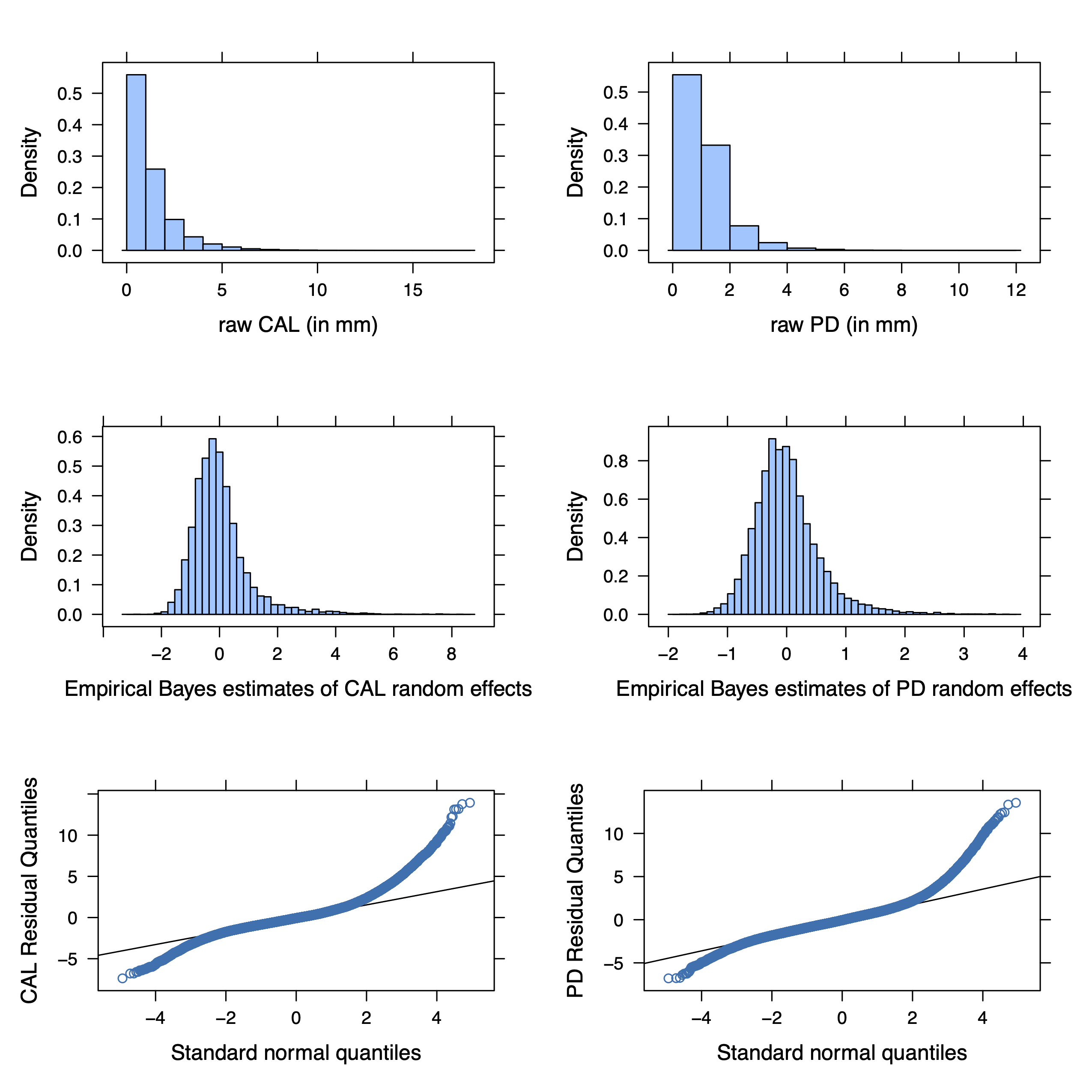}
	\caption{\label{fig:real_data_analysis_LMM}NHANES data: Plots of the density histogram of the raw PD and CAL responses (top panel), empirical Bayes' estimates of corresponding random effects (middle panel), and the Q-Q plots of corresponding standardized residuals (bottom panel), obtained after fitting linear mixed models with the Gaussian assumption to the PD and CAL responses, separately, controlling for all covariates.}
\end{figure}

To highlight the skewed and heavy-tailed nature of periodontal disease responses and other challenges in periodontal disease studies, we conduct exploratory data analysis and present results in Figures \ref{fig:real_data_analysis_LMM}, S-1 and S-2. In this paper, the prefix “S-” represents figures and tables from the online supplementary material.

PD and CAL are two commonly used biomarkers to quantify periodontal disease. CAL assesses the loss of periodontal tissue support in periodontitis, while PD indicates the depth of the periodontal pockets around teeth, both serving as critical indicators of periodontal health. We first present the histogram of raw PD and CAL responses in the top panel of Figure \ref{fig:real_data_analysis_LMM}. It is evident that both PD and CAL exhibit right skewness. Second, using the \texttt{lmer} function in the \texttt{lme4} package for \textsf{R}, we fit the classic linear mixed model with Gaussian assumptions on the random effects and residual term to the NHANES data with CAL as the response variable. Third, we separately fit the classic linear mixed model to the same data with PD as the response variable. Fourth, we present the histograms of empirical Bayes estimates of the random effects, which are the posterior means of the random effect terms, obtained using the \texttt{ranef} function in the \texttt{lme4} package, from both fitted models in the middle panel of Figure \ref{fig:real_data_analysis_LMM}. Both histograms of the random effects from the two models show right skewness, motivating us to consider an alternative to the Gaussian assumption on the random effects, opting for a more flexible choice that can accommodate skewed random effects. Finally, we present the Q-Q plots of standardized model residuals in the bottom panel of the same figure. From the Q-Q plots, the points deviate from the reference line at both the lower and upper ends for both PD and CAL residuals. Specifically, the points at the left end are below the reference line, and those at the right end are above the reference line. The deviation from the reference line indicates that the residuals have more extreme values than the Gaussian distribution, suggesting the need for a distribution with heavier tails instead.

To illustrate the prevalent non-linearity in periodontal disease studies, we applied the local polynomial regression model (LOESS) to NHANES data, using age as the sole covariate and PD and CAL as the response variables. The results are shown in Figure S-1. As depicted in the figure, both PD and CAL values generally increase with age, revealing a non-linear relationship. Notably, the increase in CAL is more marked than that of PD. From ages 30 to 50, both biomarkers show a gradual rise. However, post age 50, there is a sharper, non-linear increase in CAL, indicating a more rapid progression of periodontal disease. The classic linear mixed model assumes a linear relationship between the covariates in the fixed effect term and the response variable. Our analysis demonstrates that this assumption may not explain the relationship between risk factors and PD/CAL. While advanced machine learning algorithms are more faithful to the data-generating process, they often lack the ability to produce meaningful statistical inferences about individual risk factors. Therefore, there is a need for flexible parametric or semi-parametric models. These models can balance interpretability with the ability to handle non-linearity and non-Gaussian data distributions, providing more reliable and insightful results for periodontal disease research.

Another essential feature in the periodontal disease study is the strong correlation between PD and CAL. We present the PD and CAL scatter plot in Figure S-2. From this figure, it is evident that PD and CAL are highly correlated. We calculated the Pearson correlation coefficient between PD and CAL, which is 0.68. We also conducted the Pearson correlation test with the null hypothesis that the true correlation equals 0. The Pearson correlation test's associated $ p $-value is near 0, confirming the significant correlation between PD and CAL. The high correlation between PD and CAL motivates us to propose a model incorporating this correlation. This dual-response approach provides a more comprehensive understanding of periodontal diseases, allowing for more nuanced interpretations of disease progression and its relationship with various risk factors.

During our exploratory data analysis, we highlighted the non-Gaussian nature of periodontal diseases, the non-linear relationship between risk factors and PD/CAL, and the strong correlation between PD and CAL. While not explicitly addressed in our exploratory data analysis, it is important to note that NHANES, like other large-scale surveys, incorporates sampling weights to address the unequal probabilities of response and selection inherent in complex survey sampling methods. Numerous studies have proven that disregarding survey weight information can lead to inaccurate and unreliable inference outcomes \citep{skinner2012weighting,dong2014combining,Gunawan2020}. Moreover, NHANES offers an extensive list of risk factors for periodontal disease, underscoring the need for variable selection methods to identify the most relevant predictors from a moderately high-dimensional space. As far as we are aware, the ST-GP model proposed in this paper is the only model capable of accommodating the non-Gaussian nature of periodontal diseases, capturing the non-linear relationship between risk factors and PD/CAL, accounting for the strong correlation between PD and CAL, incorporating survey weight information, and employing a variable selection method.

The structure of the remainder of this paper is as follows: In Section \ref{sec:methodology}, we propose the ST-GP model and explain the methodology for incorporating survey weight information. In Section \ref{sec:application}, we present the formal analysis results of the NHANES data, further motivating the ST-GP model. In Section \ref{sec:simulation_studies}, we design three simulation studies demonstrating the promising performance of the proposed Gibbs sampler, of the grouped horseshoe prior, and of the adopted PBS algorithm. In Section \ref{sec:conclusion}, we conclude the paper with some remarks about the ST-GP model and several directions for future research.

\section{Methodology}\label{sec:methodology}

In this section, we propose a single index model with skewed random effects and heavy-tailed residuals. We refer to this model as the ST-GP model because the random effects and residuals jointly follow the ST distribution, and we apply the constrained Gaussian process (GP) prior from \citet{Maatouk2017} on the index function. 

Let $\mathbf{Y}_{i}^{P} = \left(Y^{P}_{i,1},Y^{P}_{i,2},\dots,Y^{P}_{i,n_i}\right)^{\top}$ and $\mathbf{Y}_{i}^{C} = \left(Y^{C}_{i,1},Y^{C}_{i,2},\dots,Y^{C}_{i,n_i}\right)^{\top}$ be the measurements of PD and CAL (in millimeter) for subject $i = 1,\dots, N$. Here $n_i$ denotes the number of teeth accounted for within the mouth for $i$-th subject. At the subject level, we propose a single index model with skewed random effects and heavy-tailed residuals as:
\begin{equation}
	\mathbf{Y}_i=\left(\begin{array}{c}
		\mathbf{Y}_i^P \\
		\mathbf{Y}_i^C
	\end{array}\right)=\left(\begin{array}{c}
		g\left(\mathbf{X}_i \boldsymbol{\beta}\right) \\
		a \times g\left(\mathbf{X}_i \boldsymbol{\beta}\right)
	\end{array}\right)+\left(\begin{array}{c}
		\boldsymbol{1}_{n_i} \\
		\boldsymbol{1}_{n_i}
	\end{array}\right) b_i+\left(\begin{array}{c}
		\boldsymbol{\epsilon}_i^P \\
		\boldsymbol{\epsilon}_i^C
	\end{array}\right),
	\label{eq:ST-GP_model}
\end{equation}
with 
\begin{equation*}
	g\left(\mathbf{X}_i \boldsymbol{\beta}\right)=\left(\begin{array}{c}
		g^{\star}\left(\mathbf{X}_i^{(1)} \boldsymbol{\beta}\right) \\
		\vdots \\
		g^{\star}\left(\mathbf{X}_i^{\left(n_i\right)} \boldsymbol{\beta}\right)
	\end{array}\right),
\end{equation*}
where $\mathbf{X}_i^{(1)}$ and $\mathbf{X}_i^{\left(n_i\right)}$ represent the first and last row of $\mathbf{X}_i$, respectively. The slope parameter $a \in (-\infty,\infty)$ differentiates the fixed effects between PD and CAL, motivated by their observed correlation. The function $g^{\star}(\cdot)$ is assumed to be a continuous monotonic increasing function on its support $[-1,1]$, with the constraint that $ g^{\star}(-1) = 0 $. For the identifiability concern, the $L_2$ norm of $\boldsymbol{\beta}$ must be 1.  As the support of $g^{\star}(\cdot)$ is defined $[-1,1]$, one need to scale $\mathbf{X}_i$ such that each row of $\mathbf{X}_i$ has $L_2$ norm no larger than 1.

The distributional assumption for the random effects and errors is expressed as follows:
\begin{equation}
	\left(\begin{array}{c}
		b_i \\
		\boldsymbol{\epsilon}_i
	\end{array}\right) \sim \operatorname{ST}_{2n_i+1}\left[\left(\begin{array}{c}
		h(\nu) \delta \\
		\boldsymbol{0}_{2n_i}
	\end{array}\right),\left(\begin{array}{cc}
		d^2 & \boldsymbol{0}_{2n_i}^{\top} \\
		\boldsymbol{0}_{2n_i} & \sigma^2 \mathbf{I}_{2n_i}
	\end{array}\right),\left(\begin{array}{c}
		\delta \\
		\boldsymbol{0}_{2n_i}
	\end{array}\right), \nu\right],
	\label{eq:single_index_model_distributional_assumption}
\end{equation}
where
$$\boldsymbol{\epsilon}_{i} = 
\left(\begin{array}{l}
	\boldsymbol{\epsilon}_i^P \\
	\boldsymbol{\epsilon}_i^C
\end{array}\right),$$
$$h(\nu) = -\sqrt{\nu / \pi} \Gamma \left(0.5 \nu - 0.5\right) / \Gamma\left(0.5 \nu\right),$$ 
$\Gamma\left(\cdot\right)$ represents the Gamma function, $d^2$ and $\sigma^2$ represent the conditional variance of the random effects and residuals, respectively, $\delta \in (-\infty,\infty)$ is the skewness parameter, and $\nu$ is the degree of freedom. Definitions and properties of the ST distribution are discussed in Section 1 of the supplementary material. If one applies the constrained GP prior on the index function $g$, then the model described in \eqref{eq:ST-GP_model} and \eqref{eq:single_index_model_distributional_assumption} is referred to as the ST-GP model. The definition of the constrained GP prior will be introduced in Section \ref{sec:single_index_function}.

In the following subsections, we will break down each assumption of the ST-GP model and explain why these assumptions are reasonable for studies on periodontal disease.

\subsection{Linear Mixed Models} \label{sec:linear_mixed_models}

In this section, we address the limitations of the classic linear mixed model, a celebrated method for modeling within-subject correlation often found in longitudinal data \citep{henderson1949estimation, henderson1950estimation, Harville1977, Laird1982}. Despite its popularity, the classic linear mixed model assumes that both the random effect term and the residual term follow a multivariate normal distribution. However, as highlighted in the exploratory data analysis presented in Section \ref{sec:introduction}, there is ample evidence suggesting that the Gaussian assumption may not hold for periodontal disease studies. Specifically, the random effects exhibit right-skewed distributions, which are not adequately captured by the normality assumption. This misalignment can lead to imprecise parameter estimates and reduced model performance.

The limitations of classic linear mixed models in handling non-normal data distributions underscore the need for more robust modeling approaches. After exploring various linear mixed models with skewed random effects proposed in the literature \citep{Rosa2003, Ho2010, lachos2010likelihood}, we adopt the ST linear mixed model from \citet{schumacher2021scale}. The main reason for this decision is that the expectations of the random effects and residuals in the ST linear mixed model from \citet{schumacher2021scale} are zeros. With this important feature, we establish a formal proof of the identifiability theorem discussed in Section \ref{sec: Identifiability of the Single Index Function}. Specifically, for the $i$-th subject, the ST linear mixed model is defined as:
\begin{equation}
	\mathbf{Y}_{i} = \mathbf{X}_{i} \boldsymbol{\beta}+ \mathbf{Z}_{i} \mathbf{b}_{i} + \boldsymbol{\epsilon}_{i},
	\label{eq:ST_linear_mixed_model}
\end{equation}
where 
\begin{equation}
	\left(\begin{array}{c}
		\mathbf{b}_i \\
		\boldsymbol{\epsilon}_i
	\end{array}\right) \sim \mathrm{ST}_{n_i + q}\left[\left(\begin{array}{c}
		h(\nu) \boldsymbol{\delta} \\
		\mathbf{0}_{n_i \times 1}
	\end{array}\right),\left(\begin{array}{cc}
		\mathbf{D} & \mathbf{0}_{q \times n_i} \\
		\mathbf{0}_{n_i \times q} & \boldsymbol{\Omega}_i
	\end{array}\right),\left(\begin{array}{c}
		\boldsymbol{\delta} \\
		\mathbf{0}_{n_i \times r}
	\end{array}\right), \nu\right].
	\label{eq:ST_linear_mixed_model_distributional_assumption}
\end{equation}
Note that the random effects and the residuals from different subjects are assumed to be independent. This model contains the standard linear mixed model as a special case, as implied by the property of the ST distribution that the SN and normal distributions are special cases of the ST distribution. As the degree of freedom $\nu$ approaches infinity and the skewness vector $\boldsymbol{\delta}$ becomes a vector of zeros, the linear mixed model in \eqref{eq:ST_linear_mixed_model} and \eqref{eq:ST_linear_mixed_model_distributional_assumption} is equivalent to the standard linear mixed model. In this context, it is important to note that $\mathbf{Y}_{i}$ can represent either PD or CAL, unlike in \eqref{eq:ST-GP_model} where it represents both PD and CAL.

Compared with the standard linear mixed model, the linear mixed model based on the ST distribution has several advantages. First, it is adequate for describing data with heavy-tailed noise. Based on the closure under linear transformation property of the ST distribution (see Equation (5) in the supplementary material), the marginal distribution of the residual term is a multivariate Student-$t$ distribution, which is well-known as a candidate for describing data with heavy tails. Second, the random effect term marginally follows the ST distribution with the shape vector $\boldsymbol{\delta}$ and is capable of modeling skewed, symmetric, or heavy-tailed subject-level effects, further enhancing its robustness in capturing non-Gaussian behavior in the data. 

For the periodontal disease study, we only include the subject-level random effect as there are no other obvious random effects to add to the model. Therefore, $\mathbf{b}_{i}$ becomes $b_{i}$, corresponding to the subject-level random effects. The modified single index model based on the ST distribution is given as,
\begin{equation}
	\mathbf{Y}_{i} = \mathbf{X}_{i} \boldsymbol{\beta} + \boldsymbol{1}_{n_{i}} b_{i} + \boldsymbol{\epsilon}_{i}.
	\label{eq:ST_linear_mixed_model_random_intercept}
\end{equation}

Although the ST linear mixed model is robust to outliers and capable of capturing data with skewness and heavy tails, it still assumes a linear association between the response variables and the fixed effects. This linearity assumption is not appropriate for periodontal disease studies, where non-linear relationships between covariates and response variables are evident, as shown in Section \ref{sec:introduction}. To address this, we propose using a single index function as a part of fixed effects. This approach removes the linearity assumption, allowing for a more flexible model that can capture the complex, non-linear relationships between covariates and the response variables PD/CAL.

\subsection{The Single Index Function} \label{sec:single_index_function}

The single index model summarizes the effects of the covariates within a single variable called the index \citep{hardle2004nonparametric}. 
We can easily incorporate the single index function into a mixed-effect model by replacing the linear fixed effect term with the index function:
\begin{equation*}
	\mathbf{Y}_{i} = g\left(\mathbf{X}_{i} \boldsymbol{\beta}\right)+ \boldsymbol{1}_{n_{i}} b_{i} + \boldsymbol{\epsilon}_{i},
\end{equation*}
with 
\begin{equation*}
	g\left(\mathbf{X}_i \boldsymbol{\beta}\right)=\left(\begin{array}{c}
		g^{\star}\left(\mathbf{X}_i^{(1)} \boldsymbol{\beta}\right) \\
		\vdots \\
		g^{\star}\left(\mathbf{X}_i^{\left(n_i\right)} \boldsymbol{\beta}\right)
	\end{array}\right).
\end{equation*}
Note that the domain and range of $g(\cdot)$ are sets of multidimensional vectors. Additionally, the domain and range of $g^{\star}(\cdot)$ are sets of scalars.

To ensure that our model parameters are uniquely determined, it is crucial to address the issue of identifiability. Identifiability refers to the ability to uniquely estimate the model parameters from the observed data. Without sufficient conditions for identifiability, the parameters $\boldsymbol{\beta}$ and the function $g(\cdot)$ may not be uniquely determined, leading to ambiguities in the interpretation and estimation of the model.

\subsubsection{Identifiability of the Single Index Function}\label{sec: Identifiability of the Single Index Function}

\citet{lin2007identifiability} provided sufficient conditions under which $ g(\cdot) $ and $\boldsymbol{\beta}$ are identifiable. To simplify notation, let a vector $X$ represent the transpose of a row of $\mathbf{X}_{i}$ and let $m(X) = g^{\star}\left(\boldsymbol{\beta}^{\top} X\right)$ be a function with a vector-valued input $X$ and a scalar-valued output. The sufficient conditions to ensure identifiability of the model in \eqref{eq:ST-GP_model} and \eqref{eq:single_index_model_distributional_assumption} are the following:

\begin{enumerate}[{1.}]
	\item The support of $ m(X) $ is assumed to be a bounded convex set with at least one interior point. (A1)
	\item We assume $ g^{\star}(\cdot) $ to be a continuous monotonic increasing function on its support. (A2)
	\item We assume the $L_{2}$ norm of $\boldsymbol{\beta}$ to be 1, such that $\|\boldsymbol{\beta}\| = 1$. (A3)
	\item We assume the degree of freedom $\nu$ to be an integer between 4 and 100. (A4)
\end{enumerate}

The assumption (A1) is essential for a formal proof of identifiability and is the same as Assumption 1 from \citet{lin2007identifiability}. We impose the assumption (A2) for the sake of clinical interpretation, as the index $\boldsymbol{\beta}^{\top} X$ can be utilized to rank patients according to their risk of periodontal diseases. The third assumption (A3) eliminates a unidentifiable situation that $g^{\star}\left(\boldsymbol{\beta}^{\top} X\right) = g^{\star}\left(\left(c\boldsymbol{\beta}^{\top} X\right)/c\right)$ for non-zero $c$.  Lastly, when the degree of freedom of $\nu$ is an integer from the assumption (A4), exactly the first $\nu$ moments of the ST distribution exist, implying that the degree of freedom $\nu$ is identifiable. Additionally, the assumption (A4) enables us to verify the condition under which we can apply the Cardano formula to prove that $\delta$ is identifiable \citep{chahal2006solution}, upon which we can establish that $d^2$ and $\sigma^2$ are identifiable.

Following Theorem 1 from \citet{lin2007identifiability}, we formally prove the identifiability of the model in \eqref{eq:ST-GP_model} and \eqref{eq:single_index_model_distributional_assumption} and present its proof in Section 3 of the supplementary material. We summarize the identifiability theorem in the following:
\begin{theorem}\label{fixed_effects_identifiability}
	If four assumptions (A1), (A2), (A3), and (A4) hold, then all parameters from the model in \eqref{eq:ST-GP_model} and \eqref{eq:single_index_model_distributional_assumption} are identifiable.
\end{theorem}

Although assumptions (A1), (A2), (A3), and (A4) are sufficient for proving the identifiability, more assumptions are needed for practical prior elicitation, which we will discuss next, specifically tailored to periodontal disease studies.

\subsubsection{Prior Elicitation on the Single Index Function}

Various Bayesian approaches are available for estimating a monotonic function \citep{bornkamp2009bayesian,shively2009bayesian,lin2014bayesian}. However, these methods encounter computational challenges when dealing with large sample sizes. \citet{chang2007shape} proposed a Bayesian approach utilizing Bernstein polynomials (BP), which is computationally more efficient than previously mentioned methods. Nonetheless, it suffers from unsatisfying empirical performance, as demonstrated in the simulation studies to be presented in Section \ref{sec:simulation_studies}. One reason for the unsatisfying empirical performance of the BP approach is that there only exists a sufficient but \emph{not} necessary condition for ensuring the monotonicity of the index function. In our paper, we adopt the constrained GP prior, which comes with a necessary \emph{and} sufficient condition for ensuring $g(\cdot)$ is coordinate-wise monotonic increasing \citep{Maatouk2017}.

To apply the constrained GP prior, we need to add one more assumption on $ g^{\star}(\cdot) $. The support of $ g^{\star}(\cdot) $ is restricted to $[-1, 1]$. Furthermore, grounded in our observation that utilizing a random intercept alone is adequate for the analysis of the real data, we add one more condition: $ g^{\star}(-1) = 0 $. This condition aligns with the reality of periodontal disease research, where the readings of PD and CAL must be non-negative. Therefore, assuming that the single index function is non-negative is reasonable. We summarize the assumption imposed on $ g^{\star}(x) $ as follows: $ g^{\star}(x) $ is defined as a continuous, monotonic increasing function on its support $[-1, 1]$, with the minimal value defined as $ g^{\star}(-1) = 0 $.

With these assumptions in place, we can now proceed to introduce the associated basis functions, $h_{k}\left(\cdot\right)$ and $\phi_{k}\left(\cdot\right)$, associated with the constrained GP prior. For given knots $-1 = u_{0} < u_{1} < \dots < u_{L} = 1$, continuous piecewise linear functions are defined as, for $k = 1,\dots,L,$
$$
h_{k}(x)= \begin{cases}0 & \text { if } x>u_{k+1} \text { or } x<u_{k-1} \\ 1 & \text { if } x=u_k \\ \text { linear } & \text { otherwise }\end{cases}.
$$
Taking integration of $h_{k}\left(x\right)$ on $(-1,x)$, we define $\psi_{k}\left(\cdot\right)$ as
$$
\psi_{k}(x)=\int_{-1}^x h_{k}(t) d t.
$$
Next, we define $\phi_{k}\left(\mathbf{X}_{i} \boldsymbol{\beta}\right)$ as a vector-valued function consisting of $n_{i}$ continuous piecewise linear functions:
$$
\phi_{k}\left(\mathbf{X}_{i} \boldsymbol{\beta}\right) = 
\begin{pmatrix}
	\psi_{k} \left(\mathbf{X}_{i}^{(1)} \boldsymbol{\beta}\right) \\
	\vdots \\
	\psi_{k} \left(\mathbf{X}_{i}^{(n_{i})} \boldsymbol{\beta}\right) \\
\end{pmatrix}.
$$
Finally, we can define the constrained GP prior and the index function as follows:
\begin{equation}
	g\left(\mathbf{X}_{i} \boldsymbol{\beta}\right)= \boldsymbol{\Phi} \boldsymbol{\xi},
	\label{eq:constrained_Gaussian_process_index_function}
\end{equation}
where $\boldsymbol{\Phi}$ is a $n_{i} \times (L + 1)$ matrix:
$$
\begin{aligned}
	\boldsymbol{\Phi} &= 
	\begin{pmatrix}
		\phi_{0} \left(\mathbf{X}_{i} \boldsymbol{\beta} \right) & \cdots & \phi_{L} \left(\mathbf{X}_{i} \boldsymbol{\beta} \right)
	\end{pmatrix} \\
	&=
	\begin{pmatrix}
		\psi_{0} \left(\mathbf{X}^{(1)}_{i} \boldsymbol{\beta}\right) & \cdots & \psi_{L} \left(\mathbf{X}^{(1)}_{i} \boldsymbol{\beta}\right) \\
		\vdots & \ddots & \vdots \\
		\psi_{0} \left(\mathbf{X}^{(n_{i})}_{i} \boldsymbol{\beta}\right) & \cdots & \psi_{L} \left(\mathbf{X}^{(n_{i})}_{i} \boldsymbol{\beta}\right)
	\end{pmatrix},
\end{aligned}
$$
and the random vector $\boldsymbol{\xi} = \left[\xi_{0},\dots,\xi_{L}\right]^{\top}$ is positive and follows a truncated multivariate normal distribution:
$$
\boldsymbol{\xi} \sim \mathcal{N}_{L+1}^{+} \left(\boldsymbol{0}_{L+1}, \boldsymbol{K}\right),
$$
representing the constrained GP prior on $\boldsymbol{\xi}$.

With the vector-valued input, $\mathbf{X}_{i} \boldsymbol{\beta}$, the index function $g\left(\cdot\right)$ is a function with vector-valued output. It is a collection of scalar-valued monotonic increasing functions:
$$
g\left(\mathbf{X}_{i}\boldsymbol{\beta}\right) = 
\begin{pmatrix}
	g^{\star}\left(\mathbf{X}_{i}^{(1)}\boldsymbol{\beta}\right) \\
	\vdots \\
	g^{\star}\left(\mathbf{X}_{i}^{(n_{i})}\boldsymbol{\beta}\right)
\end{pmatrix}
= 
\begin{pmatrix}
	\boldsymbol{\Phi}^{(1)} \boldsymbol{\xi}\\
	\vdots \\
	\boldsymbol{\Phi}^{(n_{i})} \boldsymbol{\xi}
\end{pmatrix},
$$
where $g^{\star} \left(\cdot\right)$ is a function with both scalar-valued input and output, and $\boldsymbol{\Phi}^{(1)}$ and $\boldsymbol{\Phi}^{(n_{i})}$ represent the first and last row of $\boldsymbol{\Phi}$, respectively.

By Proposition 2 of \citet{Maatouk2017}, setting $\boldsymbol{\xi}$ as a positive random vector is both a \emph{necessary and sufficient} condition for $g^{\star}\left(\cdot\right)$ to be a monotonic increasing function and for the index function $g(\cdot)$ in \eqref{eq:constrained_Gaussian_process_index_function} to be coordinate-wise monotonic increasing. 

The covariance matrix $\boldsymbol{K}$ is characterized by the Matérn kernel \citep{Rasmussen2004}, consisting of a scale parameter $\rho_{1}$, a range parameter $\rho_{2}$, and a smoothness parameter $\rho_{3}$, defined as follows:
$$
C(r)=\rho_{1}^2 \frac{2^{1-\rho_{3}}}{\Gamma(\rho_{3})}\left(\sqrt{2 \rho_{3}} \frac{r}{\rho_{2}}\right)^{\rho_{3}} B_{\rho_{3}}\left(\sqrt{2 {\rho_{3}}} \frac{r}{\rho_{2}}\right),
$$
where $r$ represents the distance between two measurements, $\Gamma\left(\cdot\right)$ denotes the gamma function, and $B_{\rho_3}\left(\cdot\right)$ is the modified Bessel function of the second kind. Inference about the smoothness parameter $\rho_{3}$ is challenging both theoretically and empirically \citep{zhang2004inconsistent}. In this paper, $\rho_{3}$ is set to $3/2$ due to the simplified analytic form of the modified Bessel function of the second kind\citep{chen2024identifiability}. Furthermore, following the suggestion by \citet{ray2020efficient}, we assume that the covariance matrix $\boldsymbol{K}$ is obtained from a regular grid in the interval $[-1,1]$, which matches the support of the index function. This results in $\boldsymbol{K}$ having a Toeplitz structure, for which there exists an associated efficient sampling algorithm.

\subsection{Correlated Response Variables}

As revealed in Section \ref{sec: Exploratory Data Analysis}, both biomarkers, PD and CAL, demonstrate a strong association. In our model, we want to incorporate both biomarkers in the same model, as PD and CAL present different aspects of periodontal disease development. To offer a comprehensive assessment of periodontal status at the tooth level within subjects, we stack PD and CAL as representative indicators of tooth-level periodontal status clustered within a subject. With this approach, researchers can effectively address the correlation between these two measures and leverage information across all teeth. Specifically, in \eqref{eq:ST-GP_model}, we include a slope parameter $a \in (-\infty, \infty)$ accounting for the association between PD and CAL, such that $g\left(\mathbf{X}_{i} \boldsymbol{\beta}\right)$ and $a \times g\left(\mathbf{X}_{i} \boldsymbol{\beta}\right)$ represent fixed effects for the $i$-th subject for PD and CAL, respectively.

\subsection{Variable Selection and Prior Elicitation}\label{sec:prior_Specification}

In this section, we aim to address one challenging aspect in analyzing the NHANES data: the relatively high number of risk factors. We also want to discuss the prior elicitation for unknown parameters $\left(a, \boldsymbol{\beta}, \delta, d^{2}, \sigma^{2}, \nu, \rho^{2}_{1}, \rho_{2}\right)$ in the ST-GP model. We suggest the following list of priors:
\begin{enumerate}
	
	\item We put a non-informative prior, a normal distribution with mean 0 and variance 1000, on the slope parameter $a$: 
	$$a \sim \mathcal{N}\left(0,1000\right).$$
	
	\item Recall that there is an identifiability restriction such that $||\boldsymbol{\beta}|| = 1$. To satisfy this restriction, the following transformation can be applied:
	$$
	\boldsymbol{\beta} = \frac{\Tilde{\boldsymbol{\beta}}}{||\Tilde{\boldsymbol{\beta}}||}.
	$$
	This transformation addresses the identifiability concern and allows for the use of the elliptical slice sampler \citep{murray2010elliptical}, which is a tuning-free sampler. As mentioned in Section \ref{sec:introduction}, traditional samplers used in existing Bayesian single index models often require careful tuning. In contrast, a tuning-free sampler simplifies the tuning process and enhances computational stability compared to samplers that require careful tuning.
	
	When the number of covariates is small, we suggest placing independent normal priors with mean 0 and variance 10 on each of $\Tilde{\boldsymbol{\beta}}$. Because, for any $c > 0$, $\Tilde{\boldsymbol{\beta}} / ||\Tilde{\boldsymbol{\beta}}|| = c\Tilde{\boldsymbol{\beta}} / ||c\Tilde{\boldsymbol{\beta}}||$, scaling the variance of the prior on $\Tilde{\boldsymbol{\beta}}$ does not alter the prior distribution of $\boldsymbol{\beta}$. 
	
	In the analysis of NHANES data, we initially focus on the influence of gender and diabetes on periodontal disease, along with other covariates. Let $ \tilde{\boldsymbol{\beta}} = \left\{ \tilde{\beta}_{\text{gender}}, \tilde{\beta}_{\text{diabetes}}, \tilde{\boldsymbol{\beta}}^{\star} \right\} $, where $ \tilde{\boldsymbol{\beta}}^{\star} $ represents all other covariates except gender and diabetes.
	
	Given the moderately high number of covariates in NHANES data, implying a moderately high dimension for $ \tilde{\boldsymbol{\beta}}$, it is important to use a shrinkage prior on the other covariates besides gender and diabetes. The same independent normal prior with mean 0 and variance 10 should be placed on $ \tilde{\beta}_{\text{gender}} $ and $ \tilde{\beta}_{\text{diabetes}} $. We elaborate on the construction of the grouped horseshoe prior in the next item of this list. 
	
	\item We put the grouped horseshoe prior on $\Tilde{\boldsymbol{\beta}}^{\star} = \left\{\Tilde{\beta}^{\star}_{j,k}: j \ge 1, k \ge 1\right\}$, such that for the $j$-th group and the $k$-th level,
	\begin{equation}
		\begin{aligned}
			\Tilde{\beta}^{\star}_{j,k} \mid \lambda_{j}, \tau &\sim \mathcal{N}\left(0,\lambda^{2}_{j} \tau^{2}\right), \\
			\lambda_{j} &\sim \mathcal{C}^{0,\infty} \left(0,1\right), \\
			\tau &\sim \mathcal{C}^{0,1} \left(0,1\right), \\
		\end{aligned}
		\label{eq:grouped_horseshoe_prior}
	\end{equation}
	where $\mathcal{C}^{0,\infty} \left(0,1\right)$ and $\mathcal{C}^{0,1} \left(0,1\right)$ represent the standard Cauchy distribution truncated to $\left(0,\infty\right)$ and the standard Cauchy distribution truncated to $\left(0,1\right)$ respectively.
	
	Last, for other data sets or for researchers who want to investigate different questions, it is advisable for researchers to determine the usage of the normal prior and the (grouped) horseshoe prior based on the specific requirements and characteristics of their data and research objectives.
	
	\item We assign a non-informative prior to the skewness parameter $\delta$, allowing the data to fully determine both the direction and magnitude of the skewness of the random effects: $$\delta \sim \mathcal{N}(0, 1000).$$
	
	\item We assign a commonly used non-informative and conjugate prior, a inverse Gamma distribution, on the variance of random effects, $d^{2}$:
	$$
	d^{2} \sim \mathcal{IG} \left(5,5\right),
	$$
	where $\mathcal{IG}(5, 5)$ denotes the inverse Gamma distribution with shape and scale parameters set to 5, characterized by the probability density function proportional to $x^{-5-1}\exp(-5/x)$.
	
	\item We assign the same non-informative and conjugate prior, $\mathcal{IG}(5, 5)$, on the variance of the residual term, $\sigma^{2}$:
	$$
	\sigma^{2} \sim \mathcal{IG} \left(5,5\right).
	$$
	
	\item To utilize the elliptical slice sampler, we place a log-normal prior on the degree of freedom:
	$$
	\log(\nu - 2) \sim \mathcal{N}(0, 1).
	$$
	This prior implies a lower bound such that $\nu > 2$, ensuring the existence of the first and second moments of the random effects and residuals.
	
	\item Similarly, for convenient use of the elliptical slice sampler, we assign the same log-normal prior on $\rho^{2}_{1}$ and $\rho_{2}$, two hyperparameters of the Matérn kernel: 
	$$
	\log\left(\rho^{2}_{1}\right) \sim \mathcal{N}(0, 1),
	$$
	and
	$$
	\log\left(\rho_{2}\right) \sim \mathcal{N}(0, 1).
	$$
	
\end{enumerate}

\subsection{The Gibbs Sampler}

The delicate prior elicitation from Section \ref{sec:prior_Specification} enables us to propose a tuning-free Gibbs sampler. Utilizing the stochastic representations of the ST-GP model (see the supplementary material for details), we can derive the conditional distributions of $a$, $\boldsymbol{\xi}$, $\delta$, $d^2$, and $\sigma^2$ in analytical forms. Note that the conditional distribution of the positive random vector $\boldsymbol{\xi}$ is a truncated multivariate normal distribution. Sampling from the conditional distribution of $\boldsymbol{\xi}$ can be done using the exact Hamiltonian algorithm for constrained multivariate normal distribution from \citet{pakman2014exact}, which has shown superior empirical performance. The conditional distributions of $a$, $\delta$, $d^2$, and $\sigma^2$ are common distributions, such as normal and inverse gamma distributions. For the sampling of $\boldsymbol{\Tilde{\beta}}$, $\nu$, $\rho^{2}_{1}$, and $\rho_{2}$, we utilize the elliptical slice sampler. Lastly, if one adopts the grouped horseshoe prior, two more parameters, $\lambda_{j}$ and $\tau$, associated with the grouped horseshoe prior, need to be inferred. The slice sampling scheme for $\lambda_{j}$ and $\tau$ is available in the online supplementary material of \citep{polson2014bayesian}. Notably, the sampling scheme for the conditional distributions of each parameter is exact. Hence, the sampler we propose is a Gibbs sampler. More details of the tailored Gibbs sampler can be found in Section 2 of the supplementary material.

\subsection{Adjustment for the Survey Weights}

The last challenge in analyzing the NHANES data we have not addressed is how to incorporate the information of the survey weights. The NHANES data is supplemented with sampling weights, which are designed to account for the varying probabilities of response and selection that are intrinsic to complex survey sampling methodologies. These weights align demographic characteristics with census data and compensate for selection biases. Ignoring them can result in biased estimates, underscoring their importance in statistical analyses\citep{skinner2012weighting}.

Several Bayesian methods tackle the issue of survey weights, including \citet{aitkin2008applications}, \citet{rao2010bayesian}, \citet{Si2015}, and \citet{Savitsky2016}. To incorporate survey weights into complex Bayesian models like the ST-GP model, \citet{Gunawan2020} proposed a resampling method called pseudo-representative samples (PRB). This method considers inference consistency and precision, reflected by frequentist coverage in repeated samples. We chose the PRB method to account for survey weights for these reasons.

There is a typo in \citet{Gunawan2020}'s paper, which could hinder readers' understanding of PRB. For convenience, we correct the typo and provide the PRB algorithm in Section 4 of the supplementary material, along with its associated Weighted Finite Population Bayesian Bootstrap algorithm (WFPBB) \citep{dong2014combining}.

\section{Application: NHANES data} \label{sec:application}

In our analysis of the NHANES data, we included several variables: gender, diabetes status, tooth site information (upper jaw, interproximal area, molar), age, ratio of family income to poverty, body mass index (BMI), high-density lipoprotein (HDL) cholesterol (mg/dL), total cholesterol (mg/dL), Glycohemoglobin percentage (HbA1c), blood lead (ug/dL), healthy eating index, binge drinking status (had at least 12 alcohol drinks), health insurance status, tobacco intake status, hypertension status, race, education level, and marital status. These variables align with the covariates used in previous studies \citep{chakraborty2014analysis, gay2018alcohol, almohamad2022association, eke2016risk, li2023association}.

As stated in Section \ref{sec:prior_Specification}, we aim to quantify the risk of periodontal diseases in four target groups: males with diabetes, males without diabetes, females with diabetes, and females without diabetes. We place independent normal priors with mean 0 and variance 10 on $ \tilde{\beta}_{\text{gender}} $ and $ \tilde{\beta}_{\text{diabetes}} $. For the other covariates, we employ a grouped horseshoe prior, which is suitable for handling the moderately high-dimensional nature of the NHANES data. This approach effectively shrinks the coefficients of irrelevant or less important predictors while preserving the significant ones. See Section \ref{sec:prior_Specification} for details of prior specification of other parameters.

\subsection{Data Preprocessing}

As with any large-scale survey data, the NHANES data is contaminated with missing values. In the data cleaning procedure, we initially eliminate any missing or immeasurable values in PD and CAL. Subsequently, we exclude observations lacking a subject identification code, as the absence of this code prevents us from determining which subject the data belongs to. For categorical variables, such as marital status, education level, hypertension status, health insurance status, bringe drinking, tobacco intake, and diabetes status, the missing rates are $0.065\%$, $0.121 \%$, $0.149\%$, $0.019\%$, $7.757 \%$, $7.673 \%$, and $2.744 \%$, respectively. Given the relatively low missing rates, retaining them as an additional level would result in extreme imbalance in these categorical variables. Hence, missing values in these variables are removed. Concerning missing values in continuous variables, including the ratio of family income to poverty, BMI, direct HDL-Cholesterol (mg/dL), total Cholesterol (mg/dL), Glycohemoglobin percentage (HbA1c), blood lead (ug/dL), and healthy eating index, we employ an ad-hoc multivariate imputation approach known as random forest imputation. This method is available in the \texttt{mice} package in the \textsf{R} programming language.

\subsection{Model Selection} \label{sec:model_selection}

An important feature of the NHANES data is that survey weights are provided to represent the varying probabilities from the complex survey sampling procedure. We adopted the PBS algorithm (details provided in the supplementary material) to adjust for survey weights and fitted ST-GP, SN-GP, N-GP, ST-BP, SN-BP, and N-BP models to the processed NHANES data. The ST-GP, SN-GP, and N-GP models have random effects and residuals following the ST, SN, and normal distributions, respectively, with the constrained GP prior on the single index function. Similarly, the ST-BP, SN-BP, and N-BP models follow the same distribution patterns but with the BP prior on the single index function. 

Using the PBS algorithm with a bootstrap size of 50, we ran the MCMC sampler for 20,000 iterations, which included 10,000 burn-in iterations and thinning every 10 draws, to approximate the posterior distribution of the parameters of interest. Then, we used the leave-one-out cross-validation information criterion (LOOIC) and the widely applicable information criterion (WAIC) as criteria for model selection. For both LOOIC and WAIC, lower values indicate a better fit. In Table \ref{tab:regression_diagnostics_loo_waic}, it is evident that models with the ST distributional assumption (ST-GP and ST-BP) outperform models with other distributional assumptions (SN-GP, SN-BP, N-GP, and N-BP). Both the LOOIC and WAIC values associated with the models with the ST distributional assumption are smaller than those of models with other distributional assumptions, indicating a better fit for the ST models. This supports the appropriateness of the ST distribution in capturing the characteristics of the PD and CAL data, including the heavy-tailed and skewed nature of the random effects. However, solely based on LOOIC and WAIC, it is indecisive which of the ST-GP and ST-BP models is better, as the LOOIC and WAIC values associated with ST-GP and ST-BP are quite close. 

To further evaluate these models, we present boxplots of residuals from all six models in Figure S-3. The red dashed lines represent the theoretical median value at 0, and black dots represent the sample median. For models with normal or SN assumptions (SN-GP, SN-BP, N-GP, and N-BP), those with the constrained GP prior exhibit residuals closer to zero compared to models with the BP prior. However, it remains inconclusive whether ST-GP or ST-BP provides a better fit, as the medians of residuals from both models are equally close to zero.

\begin{figure}
	\centering
	\includegraphics[width = 0.9\textwidth]{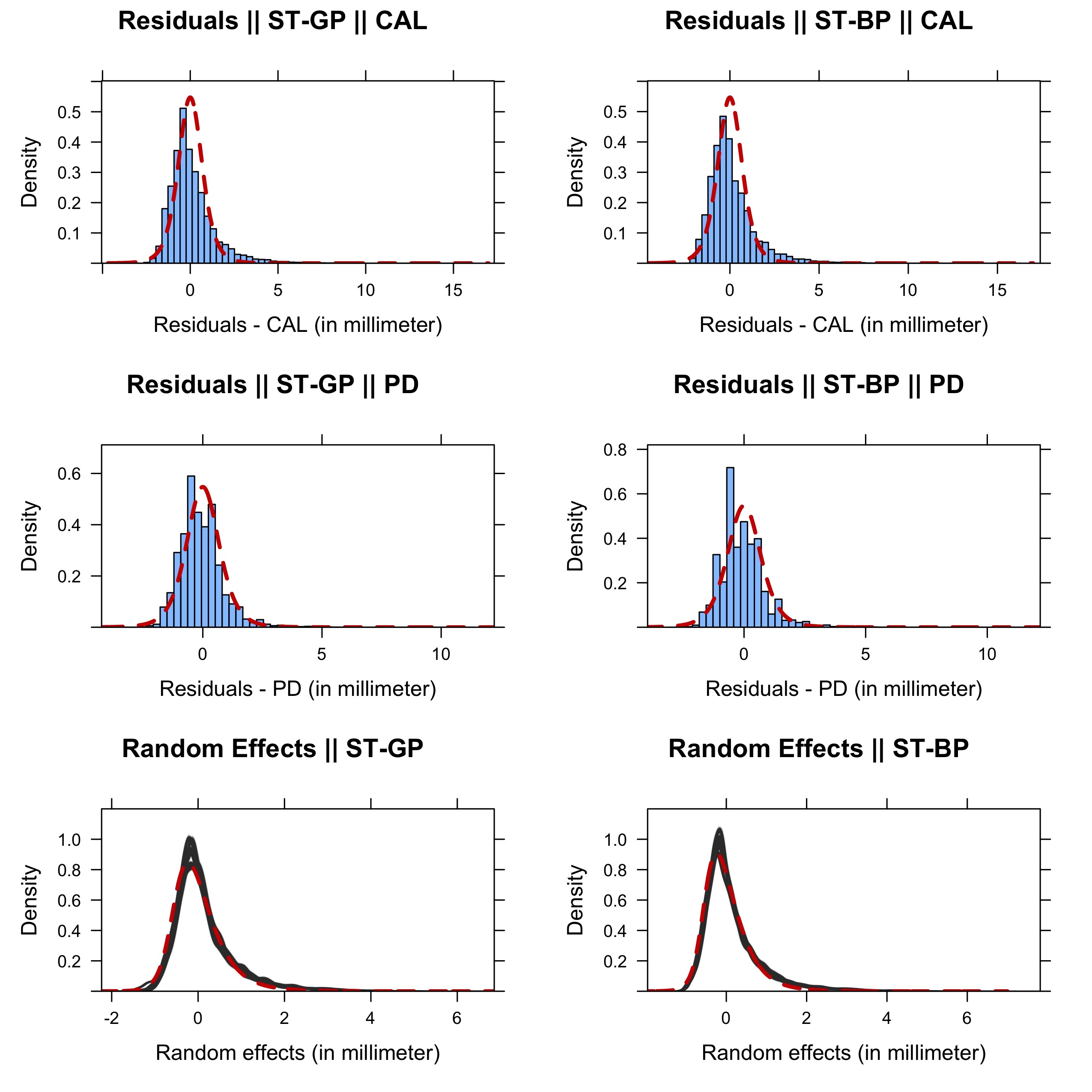}
	\caption{\label{fig:regression_diagnostics_ST_GP}NHANES data: Regression diagnostic plots from the ST-GP and ST-BP models.}
\end{figure}

Evaluating the median of residuals alone is insufficient to determine whether the ST-GP or ST-BP model provides a better fit. To further assess these models, we plot the histograms of residuals and the density curves of random effects for the ST-GP and ST-BP models in Figure \ref{fig:regression_diagnostics_ST_GP}. For both models, the residuals are expected to follow a Student-$t$ distribution. The red curves in the top and middle panels represent the density of the Student-$t$ distribution, with the estimated degrees of freedom ($\nu$) and the estimated conditional variance of residuals ($\sigma^{2}$) based on the posterior mean of these parameters. 

The residuals corresponding to CAL from both models exhibit similar histograms. However, due to a few extreme outliers around 15 millimeters, the right tails of the histograms do not align perfectly with the red curves. Despite this, the overall shapes of the histograms and the red curves for the CAL residuals match reasonably well, suggesting that the residual assumption is acceptable. In contrast, the histogram of residuals from the ST-BP model corresponding to PD shows a bar that is significantly higher than the red curve, indicating a violation of the Student-$t$ assumption for the residuals. This provides evidence that the ST-GP model is a better fit than the ST-BP model.

In addition to this empirical evidence, there is theoretical support for the superiority of the ST-GP model over the ST-BP model. The ST-GP model has a sufficient and necessary condition to ensure monotonicity, whereas the ST-BP model only has a sufficient condition to ensure monotonicity. This theoretical advantage further supports the preference for the ST-GP model in analyzing the NHANES data.

We further refine the model selection by verifying the assumption that the random effects follow a ST distribution. In the bottom panel of Figure \ref{fig:regression_diagnostics_ST_GP}, we present black curves representing the kernel density estimates of the random effects for all subjects, along with a red dashed line denoting the density of an ST distribution based on the estimated degrees of freedom ($\nu$), skewness parameter ($\delta$), and conditional variance of the random effects ($d^{2}$). The black curves closely align with the red curve, providing graphical evidence that the assumption of random effects following an ST distribution is appropriate.

We complete the model selection by presenting the inference results for the slope parameter ($a$), the skewness parameter ($\delta$), and the degrees of freedom ($\nu$) from all six models in Table \ref{tab:skewness_dof_parameters}. Notably, the 95\% credible intervals for $a$—calculated using the $2.5\%$ and $97.5\%$ quantiles of the posterior draws—exclude 0 in all six models, and the posterior means of $a$ are consistently close to 1. This confirms a strong association between PD and CAL in the real data. Additionally, the point estimates of the skewness parameter $\delta$ are all positive, and their 95\% credible intervals exclude 0, confirming the right-skewed nature of PD and CAL. Furthermore, the point estimates of the degrees of freedom ($\nu$) range between 5 and 6, with the 95\% credible intervals having an upper bound below 9, indicating the heavy-tailed nature of the real data. This comprehensive examination of the model parameters further supports the appropriateness of our modeling framework and reinforces the robustness of the ST-GP model for analyzing the NHANES data.

Based on these findings, we conclude that the ST-GP model is the best-performing model among the six models tested for the processed NHANES data.

\begin{table}
	\caption{\label{tab:regression_diagnostics_loo_waic}NHANES data: Model selection criteria (lower values indicate better fit) for the ST-GP, SN-GP, N-GP, ST-BP, SN-BP, and N-BP models. Values outside the parentheses represent the model selection criteria, with lower values indicating better model fit. Values inside the parentheses represent the percentage of the model selection criteria compared with the baseline model, the N-GP model.}
	\centering
	\begin{tabular}{llrrr}
		\toprule
		Criteria & Prior & ST & SN & N\\
		\midrule
		& GP & 9486061(89.22\%) & 10628779(99.96\%) & 10632757(100.00\%)\\
		\cmidrule{2-5}
		\multirow{-2}{*}[0.5\dimexpr\aboverulesep+\belowrulesep+\cmidrulewidth]{\raggedright\arraybackslash LOOIC} & BP & 9485727(89.21\%) & 10624356(99.92\%) & 10616629(99.85\%)\\
		\cmidrule{1-5}
		& GP & 145464603(91.35\%) & 159243394(100.00\%) & 159236668(100.00\%)\\
		\cmidrule{2-5}
		\multirow{-2}{*}[0.5\dimexpr\aboverulesep+\belowrulesep+\cmidrulewidth]{\raggedright\arraybackslash WAIC} & BP & 145191729(91.18\%) & 159112202(99.92\%) & 158805488(99.73\%)\\
		\bottomrule
	\end{tabular}
\end{table}

\begin{table}
	\caption{\label{tab:skewness_dof_parameters}NHANES data: Inference results for the slope parameter $a$, the skewness parameter $\delta$, and the degree of freedom $\nu$ from the ST-GP, SN-GP, N-GP, ST-BP, SN-BP, and N-BP models.  Numbers outside the parentheses represent the posterior mean, while numbers inside the parentheses represent the 95\% credible intervals. NA stands for “not available”.}
	\centering
	\resizebox{\columnwidth}{!}{
		\begin{tabular}{lllllll}
			\toprule
			& ST-GP & SN-GP & N-GP & ST-BP & SN-BP & N-BP\\
			\midrule
			$a$(slope) & 1.01(1.00, 1.02) & 0.99(0.97, 1.00) & 0.98(0.97, 1.00) & 1.01(1.00, 1.02) & 0.98(0.97, 1.00) & 0.97(0.96, 1.00)\\
			$\delta$(skewness) & 0.60(0.53, 0.76) & 0.78(0.70, 0.86) & NA & 0.59(0.54, 0.64) & 0.75(0.65, 0.84) & NA\\
			$\nu$(the degree of freedom) & 5.84(3.62, 8.87) & NA & NA & 5.86(3.65, 8.86) & NA & NA\\
			\bottomrule
		\end{tabular}
	}
\end{table}

\subsection{Regression Diagnostics of the ST-GP Model}

In Section \ref{sec:model_selection}, we established that the ST-GP model is the best choice for analyzing the NHANES data compared to the other five models. In this section, we provide a detailed examination of the regression diagnostics for the ST-GP model.

We begin the regression diagnostics by presenting an overview of the estimated coefficients for the covariates. Table \ref{tab:all_important_parameters} displays the point estimates (posterior means) and 95\% credible intervals for the coefficients of all covariates ($\boldsymbol{\beta}$) from the ST-GP, SN-GP, and N-GP models. Both the ST-GP and N-GP models identify three statistically significant factors influencing PD/CAL readings: whether the measurement location belongs to the upper jaw, is in the interproximal area, or is a molar. In contrast, the SN-GP model excludes the variable indicating whether the measurement location is in the upper jaw as a significant covariate, while still identifying the interproximal area and molar as statistically significant factors. Notably, the 95\% credible interval for the upper jaw variable in the SN-GP model is (-0.029, 0.000), which is on the borderline of significance. In summary, the table of estimated covariate coefficients demonstrates that models with three different likelihood assumptions—ST, SN, and normal—yield nearly identical selections of significant covariates. Other covariates do not appear to be statistically significant. This consistency across models with different distributional assumptions strengthens confidence in the identified covariates as meaningful predictors of PD/CAL readings.

\begin{figure}
	\centering
	\includegraphics[width = 0.85\textwidth]{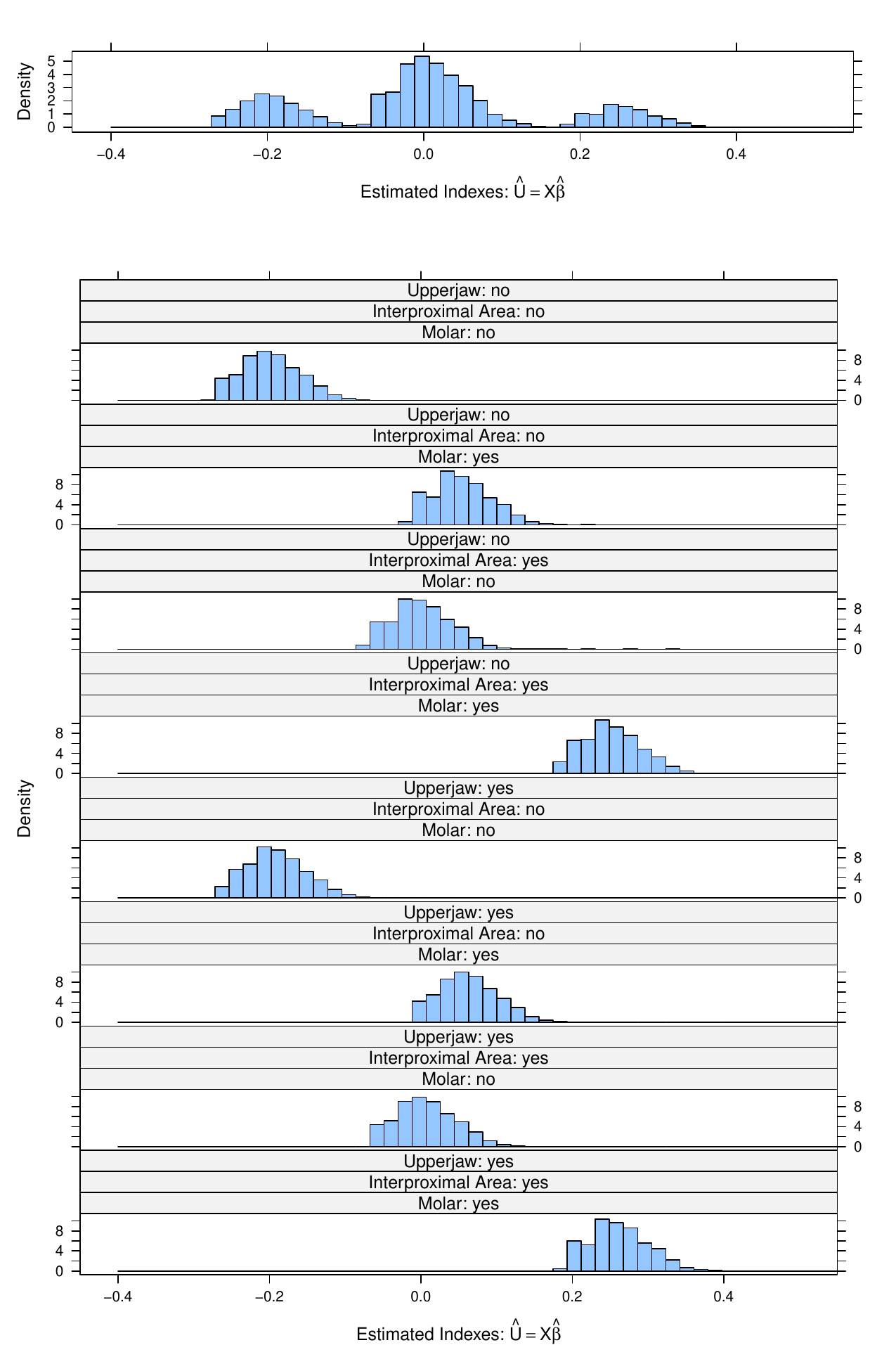}
	\caption{\label{fig:estimated_indexes}NHANES data: Histograms of estimated indexes. The X-axis represents the estimated indexes. The Y-axis represents densities.}
\end{figure}

We further investigate whether the three identified covariates are meaningful predictors of PD/CAL readings by examining the histogram of the estimated indexes in Figure \ref{fig:estimated_indexes}. In the top panel, we present the histogram of the estimated indexes, where the X-axis represents the values of the estimated indexes and the Y-axis represents the density. The estimated indexes clearly exhibit three distinct clusters: one centered around -0.2, another around 0.0, and a third around 0.25. 

To investigate the origin of the observed clusters in the estimated indexes, we stratified the data by the three covariates: upper jaw, interproximal area, and molar. The bottom panel of Figure \ref{fig:estimated_indexes} presents the histograms of the estimated indexes for each combination of these covariates. The analysis reveals distinct patterns in the distribution of estimated indexes: 1. Measurements from non-interproximal areas and non-molar sites are predominantly clustered around -0.2. 2. Measurements from interproximal areas and molar sites are primarily clustered around 0.25. 3. Other combinations of covariates result in estimated indexes clustered around 0.0.

Furthermore, molar sites are associated with higher estimated indexes compared to non-molar sites, and interproximal areas are associated with higher estimated indexes compared to non-interproximal areas. In contrast, the effect of the upper jaw covariate is less pronounced and not visually distinct in the histograms.

These findings are consistent with the inference results presented in Table \ref{tab:all_important_parameters}, where the point estimates for non-upper jaw, non-interproximal area, and non-molar are all negative. Specifically, the coefficient for non-upper jaw is -0.018, while the coefficients for non-interproximal area and non-molar are -0.469 and -0.553, respectively. This indicates that, although non-upper jaw is statistically significant, its influence is relatively modest compared to the other two covariates. 

Recall that our initial research aim was to quantify the risk of periodontal diseases across four target groups defined by the combination of gender and diabetes status. According to the covariate coefficient results in Table \ref{tab:all_important_parameters}, neither gender nor diabetes status is a statistically significant covariate. To further validate this finding, we visually compared the estimated indexes stratified by gender and diabetes status in FigureS-5, which presents histograms of the estimated indexes for each of the four groups. All four histograms exhibit the same three-cluster pattern observed in the top panel of Figure \ref{fig:estimated_indexes}. This consistency across groups provides additional evidence that neither gender nor diabetes status significantly influences the estimated indexes. These results reinforce the conclusion that gender and diabetes status are not meaningful predictors in this context, aligning with the inference results from Table \ref{tab:all_important_parameters}.

After thoroughly examining the estimated indexes, we analyzed the estimated single index function $\hat{g}^{\star}(U)$, as presented in Figure S-4. The solid line represents the estimated single index function, which exhibits clear non-linear behavior. Specifically, the function demonstrates curvature and variations in slope across different values of the indexes, confirming the non-linear nature of the relationship in periodontal disease studies, as previously discussed in Section \ref{sec: Exploratory Data Analysis}.

The translucent blue bands in Figure S-4 depict the 95\% credible interval of the single index function. Notably, the width of the credible interval varies across different values of the indexes. This variability is expected, as the estimated indexes are not uniformly distributed and instead form three distinct clusters, as discussed earlier. The non-uniform distribution of the indexes contributes to the heterogeneity in the precision of the estimated single index function across its domain.

To facilitate the practical application of our findings, we employed the variable selection approach proposed by \citet{Li2017}, which utilizes continuous shrinkage priors to identify important covariates. This approach allowed us to refine the single index formula by retaining only the most influential covariates, thereby simplifying its use for clinicians. The complete single index formula, provided in (19) in Section 5 of the supplementary material, is comprehensive but may be cumbersome for routine clinical use. To address this, we derived a concise version of the single index formula, presented in \eqref{eq:real_data_analysis_single_index_formula_simplifed}, which is more convenient for clinicians to calculate and interpret. By the monotonic increasing assumption on the single index function $g\left(U\right)$, a higher index value corresponds to a greater risk of periodontal diseases. This concise formula enables clinicians to efficiently rank patients based on their periodontal disease risk, enhancing the practical utility of our model in clinical settings.

\begin{table}
	\caption{\label{tab:all_important_parameters}NHANES data: Inference results from the ST-GP, SN-GP, and N-GP models. “ref:” represents the reference level. Numbers outside the parentheses represent the posterior mean, while numbers inside the parentheses represent the 95\% credible intervals. The 95\% credible intervals of $\boldsymbol{\beta}$ that do not contain 0 are highlighted in red.}
	\centering
	\resizebox{\columnwidth}{!}{
		\begin{tabular}[t]{lrrr}
			\toprule
			& ST-GP & SN-GP & N-GP\\
			\midrule
			Gender:female (ref:male) & -0.102(-0.367, 0.376) & -0.154(-0.475, 0.029) & -0.222(-0.631, 0.017)\\
			Diabetes:no (ref:yes) & -0.074(-0.893, 0.949) & -0.142(-0.927, 0.146) & -0.137(-0.910, 0.356)\\
			Upperjaw:no (ref:yes) & \textcolor{red}{-0.018(-0.034, -0.006)} & -0.014(-0.029, 0.000) & \textcolor{red}{-0.012(-0.020, -0.003)}\\
			Interproximal Area:no (ref:yes) & \textcolor{red}{-0.469(-0.621, -0.199)} & \textcolor{red}{-0.489(-0.603, -0.206)} & \textcolor{red}{-0.407(-0.536, -0.226)}\\
			Molar:no (ref:yes) & \textcolor{red}{-0.553(-0.704, -0.235)} & \textcolor{red}{-0.605(-0.786, -0.307)} & \textcolor{red}{-0.498(-0.655, -0.287)}\\
			\addlinespace
			Age & 0.006(-0.284, 0.243) & 0.101(-0.134, 0.364) & 0.185(-0.083, 0.650)\\
			Ratio of Family Income to Poverty & -0.023(-0.323, 0.189) & -0.028(-0.249, 0.189) & -0.060(-0.451, 0.124)\\
			BMI & 0.002(-0.186, 0.183) & -0.016(-0.220, 0.123) & -0.001(-0.175, 0.264)\\
			HDL Cholesterol (mg/dL) & 0.022(-0.144, 0.253) & -0.012(-0.164, 0.188) & 0.003(-0.142, 0.187)\\
			Total Cholesterol (mg/dL) & 0.007(-0.131, 0.208) & 0.011(-0.069, 0.129) & 0.003(-0.149, 0.151)\\
			\addlinespace
			Glycohemoglobin Percentage (HbA1c) & -0.018(-0.275, 0.245) & 0.037(-0.177, 0.312) & 0.036(-0.184, 0.388)\\
			Blood Lead (ug/dL) & 0.043(-0.262, 0.403) & 0.046(-0.133, 0.420) & 0.047(-0.144, 0.429)\\
			Healthy Eating Index & -0.003(-0.295, 0.218) & 0.025(-0.169, 0.212) & 0.030(-0.179, 0.234)\\
			Binge Drinking:no (ref:yes) & 0.011(-0.156, 0.293) & 0.011(-0.157, 0.216) & 0.018(-0.203, 0.198)\\
			Health Insurance:no (ref:yes) & 0.023(-0.209, 0.354) & 0.037(-0.250, 0.423) & 0.049(-0.157, 0.452)\\
			\addlinespace
			Tobacco Intake:no (ref:yes) & -0.034(-0.300, 0.195) & -0.052(-0.283, 0.167) & -0.080(-0.428, 0.201)\\
			Hypertension:no (ref:yes) & -0.006(-0.202, 0.212) & -0.013(-0.223, 0.136) & -0.025(-0.402, 0.141)\\
			Race:white (ref:other) & -0.070(-0.292, 0.138) & -0.028(-0.278, 0.149) & -0.011(-0.222, 0.335)\\
			Race:black (ref:other) & 0.026(-0.287, 0.295) & 0.031(-0.291, 0.337) & 0.046(-0.232, 0.363)\\
			Race:Hispanic (ref:other) & 0.040(-0.289, 0.315) & -0.026(-0.338, 0.245) & 0.024(-0.234, 0.430)\\
			\addlinespace
			Education:more than high school (ref:high school or less) & -0.032(-0.212, 0.188) & -0.053(-0.275, 0.050) & -0.091(-0.299, 0.081)\\
			Marital Status:married living with partner (ref:other) & 0.014(-0.245, 0.213) & 0.008(-0.271, 0.258) & -0.027(-0.270, 0.184)\\
			\bottomrule
		\end{tabular}
	}
\end{table}
{\small
	\begin{equation}
		\begin{aligned}
			\hat{U} = &- \mathbbm{1}\left(\text{Female}\right) \times \frac{1-0.508}{0.5} \times 0.102 
			- \mathbbm{1}\left(\text{Male}\right) \times \frac{0-0.508}{0.5} \times 0.102 \\
			& - \mathbbm{1}\left(\text{Diabetes:no}\right) \times \frac{1-0.886}{0.317} \times 0.074 
			- \mathbbm{1}\left(\text{Diabetes:yes}\right) \times \frac{0-0.886}{0.317} \times 0.074 \\
			& - \mathbbm{1}\left(\text{Interproximal Area:no}\right) \times \frac{1-0.335}{0.472} \times 0.469 
			- \mathbbm{1}\left(\text{Interproximal Area:yes}\right) \times \frac{0-0.335}{0.472} \times 0.469 \\
			& - \mathbbm{1}\left(\text{Molar:no}\right) \times \frac{1-0.753}{0.431} \times 0.553 
			- \mathbbm{1}\left(\text{Molar:yes}\right) \times \frac{0-0.753}{0.431} \times 0.553 \\
			& - \mathbbm{1}\left(\text{Race:white}\right) \times \frac{1-0.469}{0.499} \times 0.070 
			- \mathbbm{1}\left(\text{Race:not white}\right) \times \frac{0-0.469}{0.499} \times 0.070 \\
			& + \mathbbm{1}\left(\text{Race:black}\right) \times \frac{1-0.184}{0.387} \times 0.026 
			+ \mathbbm{1}\left(\text{Race:not black}\right) \times \frac{0-0.184}{0.387} \times 0.026 \\
			& + \mathbbm{1}\left(\text{Race:Hispanic}\right) \times \frac{1-0.237}{0.425} \times 0.040 
			+ \mathbbm{1}\left(\text{Race:not Hispanic}\right) \times \frac{0-0.237}{0.425} \times 0.040.
		\end{aligned}
		\label{eq:real_data_analysis_single_index_formula_simplifed}
	\end{equation}
}

\section{Simulation Studies} \label{sec:simulation_studies}

In this section, we describe three simulation studies with different purposes. In the first simulation study, we aim to demonstrate that the constrained GP prior exhibits better empirical performance than the BP for our proposed single index model. Additionally, we aim to show that the grouped horseshoe prior in \eqref{eq:grouped_horseshoe_prior} efficiently separates noise from signals. In the second simulation study, our goal is to illustrate that the PRS algorithm can effectively account for the underlying sampling mechanism in survey studies. In the last simulation study, we aim to demonstrate the robustness of our proposed single index model under model misspecification.

For all simulation studies, we replicate the non-uniform number of measurements observed in real data by setting $n_{i} = T + 2$, where $T$ follows a Poisson distribution with a mean of $8$. Each subject's data includes an associated $n_{i} \times 10$ design matrix $\mathbf{X}_{i}$. The first covariate conforms to a categorical distribution with two levels, designated as A and B, each assigned a probability of 0.5. To emulate the prevalence of diabetes observed in actual datasets, the second covariate follows a categorical distribution with two levels: diabetes and non-diabetes, assigned probabilities of 0.13 and 0.87, respectively. To investigate the performance of the grouped horseshoe prior, it is essential to include a categorical covariate with more than two levels. Thus, the third covariate is generated from a categorical distribution with three levels, each having an equal probability of 1/3. The fourth covariate also adheres to a categorical distribution with two levels, C and D, each with a probability of 0.5. In mirroring potential correlations present in real data, if the fourth covariate assumes level C, the fifth covariate follows a normal distribution with a mean of 1 and a variance of 1; otherwise, it follows a normal distribution with a mean of -1 and the same variance. The first five covariates are associated with non-zero coefficients, whereas the remaining three covariates have coefficients assigned values of zero. The sixth covariate follows a categorical distribution with three levels, each with an equal probability of 1/3. Similarly, the seventh covariate follows a categorical distribution with two levels, each with an equal probability of 0.5. Analogous to the fifth covariate, the eighth covariate follows a normal distribution with a mean of 1 if the seventh covariate assumes the first level and a mean of -1 if it assumes the second level, both with a variance of 1. Lastly, we standardized the design matrix to ensure that the $L_{2}$ norm of each row of all $\mathbf{X}_{i}$ is less than 1.

In all simulation studies, the true index function is given by 
$$g(U) = 5 \Phi\left(5U \mid 0,1\right),$$ where $\Phi\left(U \mid 0, 1\right)$ denotes the cumulative distribution function of the standard normal distribution. Other parameters of interest are set as follows: $a = 1.5$, $\delta = 0.6$, $d^{2} = 0.1$, $\sigma^{2} = 0.5$, $\nu = 5.89$, and $\Tilde{\boldsymbol{\beta}} = \left[1,1,1,1,1,1,0,0,0,0\right]^{\top}$ (equivalent to $\boldsymbol{\beta} \approx \left[0.41,0.41,0.41,0.41,0.41,0.41,0,0,0,0\right]^{\top}$). All priors are as described in Subsection \ref{sec:prior_Specification}, specifically, we put the normal prior with variance 10 on $\Tilde{\beta}_{1}$ and $\Tilde{\beta}_{2}$, and the grouped horseshoe prior on the rest of $\Tilde{\boldsymbol{\beta}}$.

\subsection{Simulation 1: GP vs BP}

In the first part of the first simulation study, to highlight the difference between the constrained GP and BP priors, we generated data from the ST-GP model with a sample size of $N = 50$ once. We present the estimated index function from the model with the constrained GP prior and the BP prior in the left and right panels of Figure S-6, respectively. The blue solid lines and red dashed lines represent the estimated index function and true index function, respectively. Blue transparent bands depict the $95\%$ credible intervals, and green dots indicate the observed index values $\mathbf{X}_{i}\boldsymbol{\beta}$. It is evident that the model with the constrained GP prior estimates the index function more precisely than the model with the BP prior, as the model with the constrained GP prior has a mean square error (MSE) of 0.83, which is smaller than the MSE of 1.31 from the model with the BP prior. To calculate the MSE of the index function, we created a uniform grid with 1000 points in $[-1,1]$. Then, at each point from the uniform grid, we calculated the difference between the estimated index function and the true index function and used this difference to calculate the MSE. Notably, the model with the constrained GP prior has a narrower credible interval than the model with the BP prior. Then, in the same simulation study, we demonstrate the effect of the grouped horseshoe prior and present the traceplots and density plots using samples from the MCMC sampler in FigureS-7. Compared with traceplots of $\beta_{1} \sim \beta_{6}$, traceplots of $\beta_{7} \sim \beta_{10}$ indicate less variance. The density plots of $\beta_{7} \sim \beta_{10}$ also indicate the shrinkage effects from the grouped horseshoe prior, as the density plots have sharp peaks at zeros, the true values of $\beta_{7} \sim \beta_{10}$.

In the second part of the first simulation study, we repeat the same simulation $100$ times with three different sample sizes for $N = 50, 100,$ and $200$ subjects. The inference results about all parameters in the fixed effect term are presented in Table S-1. Across 100 Monte Carlo replicates, we use the posterior mean as the point estimation, calculate the average bias (standard deviation in parentheses), and calculate the average MSE of the index function (standard deviation in parentheses), presenting them in the same table. Based on Table S-1, for both models with GP prior or BP prior, the largest absolute average bias of $a$ and $\boldsymbol{\beta}$ is no larger than 0.02. With the increase of sample sizes, we notice the shrinkage of bias and of the standard deviation of bias. Within the same sample size, the model with the constrained GP prior has smaller MSE of the index function than the model with BP prior. With the increase of sample size, the supremacy of the model with the constrained GP prior persists compared with the model with the BP prior, with respect to the average / standard deviation of MSE of the index function. This is expected as the constrained GP prior has a necessary and sufficient condition ensuring the monotonicity of the index function, while the BP prior has a sufficient but not necessary condition ensuring the monotonicity, as already stated in Section \ref{sec:single_index_function}. Finally, we present the bias of all parameters, excluding those in the fixed effects, in FigureS-8. It is evident that the bias of these parameters is close to zero for $N = 50$ and decreases further as the sample size increases. However, the bias of the degrees of freedom parameter ($\nu$) is relatively larger compared to the other parameters. This is expected, as the degrees of freedom parameter is known to be challenging to estimate accurately \citep{Lee2022}.

\subsection{Simulation 2: Grouped Variable Selection} \label{sec:simulation2}

In the second simulation study, we introduce a selection variable $Z$. When a sample is taken from the population, the $Z$-value for a subject in the population determines the probability of selecting that subject into the sample. Specifically, we assume that for each subject, the joint distribution of the selection variable $Z$ and the response variable $\mathbf{Y}$ is 
\begin{equation}
	\label{eq:simu1_density}
	\begin{pmatrix}
		\mathbf{Y}_{i} \\
		Z_{i}
	\end{pmatrix} \sim 
	ST_{2n_{i}+1}\left[\left(\begin{array}{c}
		\boldsymbol{\theta}_{i}+b \delta \mathbf{1}_{2 n_{i}} \\
		\mu_{z}
	\end{array}\right),\left(\begin{array}{cc}
		\boldsymbol{\Psi}_{i} & \rho \times \boldsymbol{1}_{2n_{i}} \\
		\rho \times \boldsymbol{1}_{2n_{i}}^{\top} & \sigma^{2}_{z}
	\end{array}\right),\left(\begin{array}{c}
		\delta \boldsymbol{1}_{2n_{i}}\\
		0
	\end{array}\right), \nu\right],
\end{equation}
where 
$$
\boldsymbol{\theta}_{i} = 
\left(\begin{array}{c}
	g\left(\mathbf{X}_{i} \boldsymbol{\beta}\right) \\
	a \times g\left(\mathbf{X}_{i} \boldsymbol{\beta}\right)
\end{array}\right),
$$
$g(\cdot)$ is the same index function introduced in the first simulation study,
and
$$
\boldsymbol{\Psi}_{i} = d^2 \boldsymbol{1}_{2n_{i}} \boldsymbol{1}^{\top}_{2n_{i}} + \sigma^2 \mathbf{I}_{2n_{i}}.
$$ 
Marginally, $\mathbf{Y}_{i}$ comes from the ST-GP model defined in \eqref{eq:ST-GP_model} and \eqref{eq:single_index_model_distributional_assumption}. Additionally, to replicate the intrinsic sampling mechanism in real data, we introduce a sampling mechanism here. We assume that $\mathbf{Y}_{i}$ is selected into the sample if and only if $I_{i}  = 1$, where
\begin{equation}
	\mathbb{P} \left(I_{i} = 1 \mid \mathbf{Y}_{i}, Z_{i}\right) = \mathbb{P} \left(I_{i} = 1 \mid Z_{i}\right) = \pi_{i} = \operatorname{logistic}\left(\zeta_{0} + \zeta_{1} Z_{i}\right),
	\label{eq:selection_probability}
\end{equation}
with $\operatorname{logistic}\left(\cdot\right)$ denoting the standard logistic function.

The rest of the simulation setup is as follows: We generate $N = 1000$ and $N = 2000$ (the population sample size) values of $\left(\mathbf{Y},Z\right)$ following the joint density described in Equation \eqref{eq:simu1_density}. These values are generated as a finite population, with $\mu_{z} = 0$, $\sigma^{2}_{z} = 0.6$, and $\rho = 36$. The parameters used inside the standard logistic functions are $\zeta_{0} = -1.8$ and $\zeta_{1} = 0.1$. The values of $\zeta_{1}$ and $\zeta_{0}$ control the proportion of the population selected as samples. Empirically, the selection rate is approximately $18\%$, meaning that approximately $18\%$ of the simulated finite population is selected as a sample. The rest of the simulation setting aligns with that of the first simulation study.

In Figure S-9, we present the results from the second simulation study. In the top left panel, we have the boxplot of bias of $\boldsymbol{\beta}$ across all Monte Carlo replicates with the population size 1000 and with adjustment for the survey weights information using the PRB algorithm. The bias is defined as the posterior mean of $\boldsymbol{\beta}$ minus the true values of $\boldsymbol{\beta}$. It is evident that the PRB method with a bootstrap size of 50 is adequate for adjusting for the influence of the sampling mechanism, as the largest absolute bias is no larger than 0.02 in this setting. Using the same simulation setting, we fitted the same ST-GP model to the same simulated data without adjusting for the survey weights and present the boxplots of bias in the top right panel. We refer to the inference method without adjusting for the survey weights as the naive method. From the top right panel, it is evident that failing to account for the survey weights will lead to biased estimation of $\boldsymbol{\beta}$, which is essential for the ST-GP model. Specifically, the naive method results in overestimation of $\beta_{1}, \beta_{2}, \beta_{3}$, and $\beta_{4}$ associated with three independent covariates and underestimation of $\beta_{5}$ and $\beta_{6}$ associated with two dependent covariates. We observe that the grouped horseshoe prior can separate signal from noise with both the PRB method and the naive method, as the point estimations of $\beta_{7}, \beta_{8}, \beta_{9}$, and $\beta_{10}$ are very close to their true values as zeros.

In the bottom panels, we increase the population size from 1000 to 2000 and present boxplots there. With the increase of population size, the associated standard errors of point estimations decrease in both the PRB method and the naive method. For the PRB method, the bias also decreases with the increase of population size. However, for the naive method, with the increase of population size, the bias persists indicating that the naive method leads to inconsistent estimation of $\boldsymbol{\beta}$. 

\subsection{Simulation 3: Robustness}

In the third simulation study, we generate data from the following hierarchical model:
$$
\begin{aligned}
	\mathbf{Y}^{\star}_{i} \mid b_{i} &\sim \operatorname{Laplace}_{2 n_{i} + 1} 
	\left[
	\left(
	\begin{array}{c}
		\boldsymbol{\theta}_{i} + \mathbf{1}_{2n_{i}} b_{i}\\
		\mu_{z}
	\end{array}
	\right),
	\left(
	\begin{array}{cc}
		\sigma^{2} \mathbf{I}_{2 n_{i}} & \rho \times \mathbf{1}_{2n_{i}} \\
		\rho \times \mathbf{1}^{\top}_{2n_{i}} & \sigma^{2}_{z}
	\end{array}
	\right)
	\right] \\
	b_{i} &\sim \operatorname{Gamma} \left(1,1\right),
\end{aligned}
$$
where $\mathbf{Y}^{\star}_{i} = \left[\mathbf{Y}^{\top}_{i},Z_{i}\right]^{\top}$. 

We generated data from this hierarchical model with two population sample sizes: $N = 1000$ and $N = 2000$ respectively. Same as the second simulation study, $\mathbf{Y}_{i}$ is selected into the sample if and only if $I_{i} = 1$, with its probability defined in \eqref{eq:selection_probability}. With $\sigma^{2} = \sigma^{2}_{z} = 0.6$, the values of $\left(a,\boldsymbol{\beta},\mu_{z},\rho,\zeta_{0},\zeta_{1}\right)$ and the generation of covariates $\mathbf{X}_{i}$ remains the same as those in the second simulation study. 

In Figure S-10, we observe the mild bias of point estimations of $\boldsymbol{\beta}$ across all Monte Carlo replicates. The mild bias in point estimations suggests that the model is reasonably robust to misspecifications. The robustness of the ST-GP model implies that it can still provide reliable parameter estimates even when certain assumptions of the model are violated. With the increase in population size, as depicted in the right panel, bias reduces, and the associated standard errors of point estimations decrease.The observed improvements in bias and standard errors with increased sample size highlight the potential applicability of the ST-GP model in real-world data with large sample sizes. Overall, the results of this simulation study reinforce the robustness of the ST-GP model.

\section{Conclusion} \label{sec:conclusion}

In this paper, we proposed the ST-GP model, a Bayesian single-index model with skewed random effects terms that follow a skew-t (ST) distribution and potential heavy-tailed noise terms that follow a Student-$t$ distribution. We utilized the PBS algorithm to incorporate the survey weights that are commonly available in large-scale survey data, such as the NHANES data. We utilized an innovative prior, the constrained GP prior, on the index function $ g(U) $, which is assumed to be a non-decreasing function for the sake of interpretability. The most important advantage of the constrained GP prior is that it provides a necessary and sufficient condition to ensure the monotonicity of the index function. Because both PD and CAL are popular biomarkers for quantifying periodontal diseases, we “stack” PD and CAL and introduce a slope parameter $ a $ that connects the fixed effects terms of PD and CAL. By doing so, we can include both biomarkers in the same model and quantify the association between both biomarkers and covariates. We utilized the grouped horseshoe prior, which is suitable for both continuous variables and multi-level categorical variables, for the purpose of variable selection. The number of covariates in the NHANES data is relatively large, so using a shrinkage prior becomes essential for discovering the true factors that are associated with periodontal diseases. Taking advantage of the hierarchical representation of the proposed ST-GP model, we designed a tuning-free Gibbs sampler tailored to the ST-GP model. The tuning-free Gibbs sampler is more convenient for practitioners compared to other commonly used MCMC algorithms such as the Metropolis-Hastings and Hamiltonian Monte Carlo methods.

We demonstrated that the proposed ST-GP model, with the PBS algorithm that accounts for survey weights information, is more appropriate than five other models (ST-BP, SN-GP, SN-BP, N-GP, and N-BP) for the NHANES data, with a focus on quantifying periodontal diseases. We found much evidence showing that incorporating the skewed random effects term and including the heavy tail in the noise term is necessary for the NHANES data, as demonstrated in Section \ref{sec:application}. By comparing the density plot of the estimated random effects with the theoretical distribution (ST distribution) and comparing the histogram of the residuals of both biomarkers with the theoretical density plot (Student-$t$ distribution), we concluded that the proposed ST-GP model is suitable for the NHANES data.

Before delving into the real data analysis, our initial hypothesis was that diabetes status and gender are the two most influential factors affecting periodontal diseases. We plotted the index function by the four subgroups, which are the combinations of gender and diabetes status. However, as shown in the coefficient estimation results in Table \ref{tab:all_important_parameters}, the influence of gender and diabetes is not statistically significant, and the coefficient associated with diabetes status is close to 0. However, we found a clustering pattern in the index function plot, indicating that the measuring location—whether it is an interproximal area or not—and whether the tooth being measured is a molar or not are the two most important factors in the context of periodontal disease study. This finding is verified both by plotting the index function by the four subgroups, which are the combinations of the interproximal area factor and the molar factor, and by the inference of the coefficients associated with the interproximal area and molar factors from Table \ref{tab:all_important_parameters}.

After demonstrating the applicability of the ST-GP model in the real data application in Section \ref{sec:application}, we designed three simulation studies in Section \ref{sec:simulation_studies}. In the first simulation study, we illustrated the superiority of the constrained GP prior over another commonly used prior, the BP prior, under the monotonicity assumption on the index function. Additionally, we demonstrated the effectiveness of the grouped horseshoe prior in the same simulation study. In the second simulation study, we demonstrated the necessity of incorporating survey weight information and the effectiveness of the PBS algorithm for adjusting the survey weights. In the last simulation study, we illustrated the robustness of the ST-GP model under a specific model misspecification.

A future direction for exploring the single index model with skewed random effects terms and heavy-tailed noise terms includes proposing an equivalent Frequentist single index model. One possible approach is to replace the constrained GP prior with a deep neural network model. An innovative procedure is required to incorporate survey weight information into a deep neural network model. This approach can leverage the flexibility and powerful approximation capabilities of deep learning while integrating survey weights to address the complexities of large-scale survey data, such as NHANES.

Finally, we developed an \textsc{R} package named \texttt{MSIMST}, which is publicly available on \texttt{CRAN}. This package implements the methodology proposed in this paper, including all six models discussed (ST-GP, SN-GP, N-GP, ST-BP, SN-BP, and N-BP) as well as the PBS algorithm \citep{Liu2024}. 

\section*{Declaration of generative AI in scientific writing}

During the preparation of this work the authors used generative pre-trained transformer models in order to check grammar. After using these tool/service, the authors reviewed and edited the content as needed and take full responsibility for the content of the publication.

\bibliographystyle{apalike}
\bibliography{mybibilo.bib} 

\end{document}


\maketitle

\section{Skewed Distributions}\label{supp:skewed_distributions}

In this section, we introduce the definition of the ST distribution by first explaining the construction of the SN distribution. The construction of the SN distribution begins with a linear combination of two \emph{independent} normal distributions. A random vector $\mathbf{Y}$ follows a skew-normal (SN) distribution with a $p\times 1$ location vector $\boldsymbol{\mu}$, a $p\times p$ scale matrix $\boldsymbol{\Omega}$, and a $p\times 1$ skewness vector $\boldsymbol{\delta}$, denoted as $\mathbf{Y} \sim \text{SN}_{p} \left(\boldsymbol{\mu}, \boldsymbol{\Omega}, \boldsymbol{\delta} \right)$, if it can be expressed as:
\begin{equation}
	\mathbf{Y} = \boldsymbol{\mu} + \boldsymbol{\delta} | X_{0} | + \mathbf{X}_{1},
	\label{eq:SN_stochastic_representation}
\end{equation}
where $X_{0}$ follows a univariate standard normal distribution, and $\mathbf{X}_{1}$ follows a multivariate normal distribution with zero mean and a covariance matrix $\boldsymbol{\Omega}$. The random variable that follows the truncated normal distribution, $|X_{0}|$, along with the skewness vector $\boldsymbol{\delta}$, brings skewness into the SN distribution. 

By introducing one more latent variable, denoted as $U$, which is independent of $X_{0}$ and $\mathbf{X}_{1}$ and follows a Gamma distribution with shape and rate parameters both equal to $\nu/2$, i.e., $U \sim \text{Gamma} \left(\nu/2,\nu/2\right)$, where its density function is proportional to $u^{0.5\nu - 1} \exp\left(-0.5\nu u\right)$, we can construct the ST distribution as follows:
\begin{equation}
	\mathbf{Y} = \boldsymbol{\mu} + U^{-1/2} \left(\boldsymbol{\delta} | X_{0} | + \mathbf{X}_{1}\right),
	\label{eq:ST_stochastic_representation}
\end{equation}
which is denoted as $\mathbf{Y} \sim \text{ST}_{p} \left(\boldsymbol{\mu},\boldsymbol{\Omega},\boldsymbol{\delta},\nu\right)$. Adding the new latent variable $U$ introduces heavy tail and high kurtosis features into the ST distribution.

The stochastic representations of the ST and SN distributions in \eqref{eq:ST_stochastic_representation} and \eqref{eq:SN_stochastic_representation} are not only useful for sampling from the ST/SN distributions but also imply the relationship between the ST distribution and the SN distribution. As the degree of freedom parameter $\nu$ approaches infinity, $U$ converges to $1$ in probability, and therefore, the ST distribution converges to the SN distribution. Additionally, \eqref{eq:SN_stochastic_representation} and \eqref{eq:ST_stochastic_representation} imply that the normal distribution is a special case of the SN distribution, and that both the normal distribution and the Student-$t$ distribution are special cases of the ST distribution. With the shape vector $\boldsymbol{\delta}$ set as a vector of zeros, the random vector defined in \eqref{eq:ST_stochastic_representation} follows a multivariate Student-$t$ distribution with the degree of freedom parameter $\nu$, while the random vector defined in \eqref{eq:SN_stochastic_representation} follows a multivariate normal distribution.

Finally, the stochastic representation of the ST distribution in \eqref{eq:ST_stochastic_representation} also implies an equivalent hierarchical representation:
\begin{equation}
	\begin{aligned}
		\mathbf{Y} \mid S, U &\sim \mathcal{N}_{p}\left(\boldsymbol{\mu} + u^{-1/2} s \boldsymbol{\delta}, u^{-1} \boldsymbol{\Omega}\right), \\
		S &\sim \mathcal{N}^{+} \left(0, 1\right),\\
		U &\sim \operatorname{Gamma}\left(\nu/2, \nu/2\right).
	\end{aligned}
	\label{eq:ST_hierarchical_representation}
\end{equation}
Integrating out two latent variables, $S$ and $U$, we obtain the density function of the ST distribution as:
\begin{equation}
	f_{\mathbf{Y}}(\mathbf{Y}) = 2 t_{p}(\mathbf{Y} \mid \boldsymbol{\mu}, \boldsymbol{\Sigma}, \nu) T\left(\boldsymbol{\delta}^{\top} \boldsymbol{\Sigma}^{-1}(\mathbf{Y}-\boldsymbol{\mu}) \sqrt{\frac{\nu+p}{\nu+d(\mathbf{Y})}} \mid 0, \Lambda, \nu+p\right),
	\label{eq:ST_pdf}
\end{equation}
where $t_{p}$ represents the density function of a $p$-dimensional multivariate Student-$t$ distribution, $T$ represents the cumulative distribution function of a univariate Student-$t$ distribution. Additional $\boldsymbol{\Sigma}$ and $\boldsymbol{\Lambda}$ are given as follows:
$$
\boldsymbol{\Sigma} = \boldsymbol{\Omega} + \boldsymbol{\delta} \boldsymbol{\delta}^{\top},
$$
$$
\Lambda = 1 - \boldsymbol{\delta}^{\top} \boldsymbol{\Sigma}^{-1} \boldsymbol{\delta}.
$$
Furthermore, $d(\mathbf{Y})$ is defined as:
$$
d(\mathbf{Y}) = (\mathbf{Y} - \boldsymbol{\mu})^{\top} \boldsymbol{\Sigma}^{-1} (\mathbf{Y} - \boldsymbol{\mu}).
$$
Both representations of the ST distribution in \eqref{eq:ST_stochastic_representation} and \eqref{eq:ST_hierarchical_representation}, along with its density function in \eqref{eq:ST_pdf}, are utilized in the tailored Gibbs sampler.

A notable feature of the ST distribution is its closure under linear transformation, as demonstrated in Proposition 5 of \citet{schumacher2021canonical}. That is, if $\mathbf{Y} \sim \operatorname{ST}_{p}\left(\boldsymbol{\mu},\boldsymbol{\Omega},\boldsymbol{\delta},\nu\right)$ then
\begin{equation}
	\mathbf{AY} + \mathbf{b} \sim \operatorname{ST}_{m}\left(\mathbf{A}\boldsymbol{\mu} + \mathbf{b},
	\mathbf{A}\boldsymbol{\Omega}\mathbf{A}^{\top},\mathbf{A}\boldsymbol{\delta},\nu\right),
	\label{eq:ST_linear_transformation}
\end{equation}
where $\mathbf{A}$ is a $m\times p$ matrix and $\mathbf{b}$ is a vector of length $m$. This feature is essential for constructing mixed-effect models based on the ST distribution, as discussed in Sections 2.1 and 2.2 from the main article. 

\subsection{The Hierarchical Representation}
Using Proposition 5 of \citet{schumacher2021canonical}, we have
\begin{equation}
	\mathbf{Y}_i \sim \operatorname{ST}_{2n_{i}}\left(\boldsymbol{\theta}_i+ h(\nu) \delta \boldsymbol{1}_{2n_{i}}, \boldsymbol{\Psi}_i, \delta \boldsymbol{1}_{2n_{i}}, \nu\right),
\end{equation}
where
$$
\boldsymbol{\theta}_i = 
\left(\begin{array}{c}
	g\left(\mathbf{X}_i \boldsymbol{\beta}\right) \\
	a \times g\left(\mathbf{X}_i \boldsymbol{\beta}\right)
\end{array}\right)
$$
and
$$\boldsymbol{\Psi}_i = d^2 \boldsymbol{1}_{2n_{i}} \boldsymbol{1}^{\top}_{2n_i} + \sigma^2 \mathbf{I}_{2n_i \times 2n_i}$$ 
represents a covariance matrix characterized by a compound symmetry structure, with readily available closed-form expressions for its inverse and determinant.

From Proposition 6 of \citet{schumacher2021canonical}, we have the following stochastic representation,
\begin{equation}
	\begin{aligned}
		\mathbf{Y}_i \mid \cdot &\sim \mathcal{N}_{2n_i} \left( \boldsymbol{\theta}_i + \boldsymbol{1}_{2n_{i}}b_i, u^{-1}_{i} \sigma^2 \mathbf{I}_{2n_{i}}\right) \\
		b_i \mid \cdot &\sim \mathcal{N}\left(\delta \left( h(\nu) + s_i\right), u^{-1}_{i} d^2 \right) \\
		S_i \mid \cdot &\sim \mathcal{N}^{+} \left( 0, u_{i}^{-1}\right) \\
		U_i &\sim \operatorname{Gamma} \left(\nu/2 , \nu/2 \right), \quad i = 1,\dots,N.
	\end{aligned}
	\label{eq:stochastic representation}
\end{equation}
Here, $\mathcal{N}^{+} \left(0, u_i^{-1}\right)$ represents the half-normal distribution with a location parameter of 0 and a scale parameter of $\sqrt{u_i^{-1}}$.

We utilize Equation \eqref{eq:stochastic representation} to derive updating equations for the parameters $ \delta $, $ S_i $, $ U_i $, $ \boldsymbol{\xi} $, $ b_i $, $ \sigma^2 $, and $ d^2 $. The remaining parameters, including $ \boldsymbol{\beta} $, $ \nu $, and the hyperparameters associated with the constrained GP prior, are updated using the elliptical slice sampler algorithm.

\section{The Gibbs Sampler} \label{supp:Gibbs}
To simplify notation, let 
$$g\left(\mathbf{X}_{i} \boldsymbol{\beta}\right) = g_{i}.$$
\subsection*{Update \texorpdfstring{$a$}{a}}
The prior for $a$ is
$$
a \sim \mathcal{N} \left(0, \sigma^2_{a} \right).
$$
Let
$$
\boldsymbol{\Omega}_{i,a} = u_{i}^{-1} \sigma^{2} \mathbf{I}_{n_{i}},
$$
\begin{equation}
	\mathbf{Y}_{i,a}^{\star} = \mathbf{Y}_{i}^{C} - \boldsymbol{1}_{n_i} b_{i}.
\end{equation}
After simple algebra,
\begin{equation}
	a \mid \cdot \sim \mathcal{N}\left(\frac{\sum_{i=1}^N g_{i}^{\top} \boldsymbol{\Omega}_{i, a}^{-1} \mathbf{Y}_{i, a}^{\star}}{\sigma_{a}^{-2}+\sum_{i=1}^N g_{i}^{\top} \boldsymbol{\Omega}_{i, a}^{-1} g_{i}}, \frac{1}{\sigma_{a}^{-2}+\sum_{i=1}^N g_{i}^{\top} \boldsymbol{\Omega}_{i, a}^{-1} g_{i}}\right).
\end{equation}

\subsection*{Update \texorpdfstring{$\boldsymbol{\xi}$}{xi}}

Given hyperparameters $\rho_{1}^{2}$ and $\rho_{2}$, the distribution for $\boldsymbol{\xi} = \left(\xi_0, \dots, \xi_{L}\right)^{\top}$ is
$$
\boldsymbol{\xi} \mid \cdot \sim \mathcal{N}^{+}_{L+1} \left(\boldsymbol{0}_{L + 1}, \boldsymbol{K}\right).
$$
Let 
$$
\mathbf{Y}_{i,\boldsymbol{\xi}}^{\star}=
\left(\begin{array}{c}
	\mathbf{Y}_i^P - \boldsymbol{1}_{n_i \times 1} \left( h(\nu) \delta + s_i \delta\right)\\
	\left(\mathbf{Y}_i^C - \boldsymbol{1}_{n_i \times 1} \left( h(\nu) \delta + s_i \delta\right)\right)/a
\end{array}\right),
$$
$$
\boldsymbol{\Omega}_{i,\boldsymbol{\xi}} = 
\left(\begin{array}{cc}
	\mathbf{I}_{n_i} & \mathbf{0}_{n_i} \\
	\mathbf{0}_{n_i} & a^{-1} \mathbf{I}_{n_i}
\end{array}\right) u^{-1}_i \boldsymbol{\Psi}_i\left(\begin{array}{cc}
	\mathbf{I}_{n_i} & \mathbf{0}_{n_i} \\
	\mathbf{0}_{n_i} & a^{-1} \mathbf{I}_{n_i}
\end{array}\right),
$$
and
$$
\boldsymbol{\Phi}_i=
\left(\begin{array}{c}
	\phi_0\left(\mathbf{X}_i \boldsymbol{\beta}\right), \ldots, \phi_L\left(\mathbf{X}_i \boldsymbol{\beta}\right) \\
	\phi_0\left(\mathbf{X}_i \boldsymbol{\beta}\right), \ldots, \phi_L\left(\mathbf{X}_i \boldsymbol{\beta}\right)
\end{array}\right).
$$
After simple algebra,
$$
\boldsymbol{\xi} \mid \cdot \sim \mathcal{N}_{L+1}^{+} \left( \left(\boldsymbol{K}^{-1} + \sum_{i = 1}^{N} \boldsymbol{\Phi}_i^{\top} \boldsymbol{\Omega}_{i,\boldsymbol{\xi}}^{-1} \boldsymbol{\Phi}_i\right)^{-1} \left(\sum_{i = 1}^{N} \boldsymbol{\Phi}_i^{\top} \boldsymbol{\Omega}_{i,\boldsymbol{\xi}}^{-1} \mathbf{Y}_{i, \boldsymbol{\xi}}^{\star}\right),
\left(\boldsymbol{K}^{-1} + \sum_{i = 1}^{N} \boldsymbol{\Phi}_i^{\top} \boldsymbol{\Omega}_{i,\boldsymbol{\xi}}^{-1} \boldsymbol{\Phi}_i\right)^{-1}\right).
$$

\subsection*{Update \texorpdfstring{$\delta$}{delta}}

The prior for $\delta$ is
$$
\delta \sim \mathcal{N}\left(0, \sigma^{2}_{\delta}\right).
$$
Let
$$
\mathbf{Y}^{\star}_{i,\delta} = 
\left(\begin{array}{c}
	\mathbf{Y}_i^P \\
	\mathbf{Y}_i^C
\end{array}\right) -
\left(\begin{array}{c}
	g_{i} \\
	a g_{i}
\end{array}\right),
$$
and
$$
\boldsymbol{\Omega}_{i,\delta} = u^{-1}_{i} \boldsymbol{\Psi}_{i}.
$$
After simple algebra,
$$
\delta \mid \cdot \sim \mathcal{N} \left(\frac{\sum_{i=1}^{N} \left(h(\nu) + s_i\right) \boldsymbol{1}_{2n_i}^{\top} \boldsymbol{\Omega}_{i,\delta}^{-1} \mathbf{Y}_{i, \delta}^{\star}}{\sigma^{-2}_{\delta} + \sum_{i=1}^{N} \left(h(\nu) + s_i\right)^{2}\boldsymbol{1}^{\top}_{2n_i} \boldsymbol{\Omega}_{i,\delta}^{-1} \boldsymbol{1}_{2n_i}},
\frac{1}{\sigma^{-2}_{\delta} + \sum_{i=1}^{N} \left(h(\nu) + s_i\right)^{2}\boldsymbol{1}_{2n_i}^{\top} \boldsymbol{\Omega}_{i,\delta}^{-1} \boldsymbol{1}_{2n_i}}\right).
$$

\subsection*{Update \texorpdfstring{$S_i$}{Si}}

Let 
$$
\mathbf{Y}^{\star}_{i,S_i} = \mathbf{Y}_{i} - \boldsymbol{\theta}_i - h(\nu) \delta \boldsymbol{1}_{2n_{i}},
$$
and 
$$
\boldsymbol{\Omega}_{i,S_i} = u_{i}^{-1}\boldsymbol{\Psi}_i.
$$
After simple algebra,
$$
S_{i} \mid \cdot \sim \mathcal{N}^{+} \left(\frac{\delta \boldsymbol{1}_{1\times 2n_i} \boldsymbol{\Omega}_{i}^{-1} \mathbf{Y}_{i}^{\star}}{u_{i} + \delta^2 \boldsymbol{1}_{1 \times 2 n_i} \boldsymbol{\Omega}_{i, \delta}^{-1} \boldsymbol{1}_{2 n_i \times 1}}, \frac{1}{u_{i} + \delta^2 \boldsymbol{1}_{1 \times 2 n_i} \boldsymbol{\Omega}_{i, \delta}^{-1} \boldsymbol{1}_{2 n_i \times 1}}\right).
$$

\subsection*{Update \texorpdfstring{$U_i$}{Ui}}

Let 
$$
\mathbf{Y}^{\star}_{i,U_{i}} = \mathbf{Y}_{i} - \boldsymbol{\theta}_{i} - h(\nu) \delta \boldsymbol{1}_{2n_{i} \times 1} - \delta s_i \boldsymbol{1}_{2n_{i} \times 1}.
$$
$$
U_{i} \mid \cdot \sim \operatorname{Gamma} \left(0.5\left(2n_{i} + \nu + 1\right), 0.5 \left({\mathbf{Y}^{\star}_{i,U_{i}}}^{\top} \boldsymbol{\Psi}_{i}^{-1} \mathbf{Y}^{\star}_{i,U_{i}} + s^{2}_{i} + \nu\right)\right).
$$

\subsection*{Update \texorpdfstring{$b_i$}{bi}}
Let 
$$
\mathbf{Y}^{\star}_{i,b_{i}} = \mathbf{Y}_{i} - \boldsymbol{\theta}_{i}.
$$
$$
b_{i} \mid \cdot \sim \mathcal{N}\left(\frac{\sigma^{-2}\left(\boldsymbol{1}^{\top}_{n_{i}} \mathbf{Y}^{\star}_{i,b_{i}}\right) + \delta\left(h(\nu) + s_i\right)d^{-2}}{2 n_{i}\sigma^{-2} + d^{-2}}, \frac{1}{2 n_{i}u_{i}\sigma^{-2} + u_{i}d^{-2}}\right).
$$

\subsection*{Update \texorpdfstring{$\sigma^{2}$}{sigmasq}}

The prior for $\sigma^{2}$ is
$$
\sigma^{2} \sim \text{Inverse Gamma} \left(a_{\sigma^2}, b_{\sigma^2}\right).
$$
Let 
$$
\mathbf{Y}^{\star}_{i,\sigma^2} = \mathbf{Y}_{i} - \boldsymbol{\theta}_{i} - \boldsymbol{1}_{2n_{i}} b_{i}.
$$
$$
\sigma^2 \mid \cdot \sim \text{Inverse Gamma} \left(a_{\sigma^2} + \sum_{i=1}^{N} n_{i}, b_{\sigma^2} + 0.5 \sum_{i=1}^{N} u_{i}\left({\mathbf{Y}^{\star}_{i,\sigma^2}}^{\top}\mathbf{Y}^{\star}_{i,\sigma^2}\right)\right).
$$

\subsection*{Update \texorpdfstring{$d^{2}$}{dsq}}
The prior for $d^{2}$ is
$$
d^{2} \sim \text{Inverse Gamma} \left(a_{d^2}, b_{d^2}\right).
$$
$$
d^2 \mid \cdot \sim \text{Inverse Gamma} \left(0.5N + a_{d^{2}}, b_{d^{2}} + 0.5\sum_{i=1}^{N} u_{i}\left(b_{i} - \delta\left(h(\nu) + s_{i}\right)\right)^2\right).
$$

\section{Identifiability Theorem} \label{supp:Identifiability_Theorem}

Recall a famous result regarding the moments of a Gamma distribution. Let $U \sim \operatorname{Gamma}\left(\nu/2, \nu/2\right)$, then
$$M_1 := \mathbb{E}\left(U^{-1/2}\right) = \frac{\left(\nu/2\right)^{1/2}}{\Gamma\left(\nu/2\right)} \Gamma\left(\nu/2 - 1/2\right), \quad \text{if } \nu > 1,$$
$$M_2 := \mathbb{E}\left(U^{-2/2}\right) = \frac{\left(\nu/2\right)^{2/2}}{\Gamma\left(\nu/2\right)} \Gamma\left(\nu/2 - 2/2\right) = \frac{\nu/2}{\nu/2 - 1},\quad \text{if } \nu > 2,$$
and
$$M_3 := \mathbb{E}\left(U^{-3/2}\right) = \frac{\left(\nu/2\right)^{3/2}}{\Gamma\left(\nu/2\right)} \Gamma\left(\nu/2 - 3/2\right),\quad \text{if } \nu > 3.$$

\begin{lemma} \label{skewness_identifiable}
	Let 
	\begin{equation}
		Y \sim \mathrm{ST}_{1}\left(-M_1 \sqrt{2 / \pi} \delta, \sqrt{d^2+\sigma^2}, \delta, \nu\right).
		\label{eq:identifiability_nu_delta}
	\end{equation} 
	Or equivalently, let
	$$Y \sim \mathrm{ST}_{1}\left(h\left(\nu\right) \delta, \sqrt{d^2+\sigma^2}, \delta, \nu\right).$$
	If the assumption (A4) holds, then the degree of freedom $\nu$ and the skewness parameter $\delta$ are identifiable.
\end{lemma}
\begin{proof}
	First, we want to show that the degree of freedom $\nu$ is identifiable.
	
	Let $\mathcal{M}_{1}$ and $\mathcal{M}_{2}$ represent models in \eqref{eq:identifiability_nu_delta} with the parameterizations $\left[\sigma_1, d_1, \delta_1, \nu_1\right]$ and $\left[\sigma_2, d_2, \delta_2, \nu_2\right]$, respectively. Suppose $\mathcal{M}_{1} = \mathcal{M}_{2}$, that is, $Y_{1}$ is equivalent to $Y_{2}$ in distribution, denoted as $Y_{1} \stackrel{d}{=} Y_{2}$. Specifically, $Y_{1}$ and $Y_{2}$ follow the univariate ST distribution with the parameterizations $\left[\sigma_1, d_1, \delta_1, \nu_1\right]$ and $\left[\sigma_2, d_2, \delta_2, \nu_2\right]$, respectively. From $Y_{1} \stackrel{d}{=} Y_{2}$, we know that the collection of all moments (of all orders) of $Y_{1}$ and $Y_{2}$ must be equal, when they exist. Recall the stochastic representation of the ST distribution in \ref{eq:ST_stochastic_representation}, when $\nu$ is an integer, only the first $\nu$ moments of $Y$ exist. For example, if the first four moments of $Y_{1}$ and $Y_{2}$ exist and the fifth and higher moments of them does not exit, then $\nu_1 = \nu_2 = 4$. Thus, under the assumption (A4), the degree of freedom $\nu$ is identifiable regardless the value of the skewness parameter $\delta$.
	
	Second, we want to show that the skewness parameter $\delta$ are identifiable. Under the assumption (A4), the first three moments of $Y$ exist. The first moment of $Y$ is zero. The second and third moments of $Y$ are
	\begin{equation}
		\mathbb{E}\left(Y^2\right) = C_1 \delta^2 + C_2 \left(\sigma^2 + \delta^2\right),
		\label{eq:ST_second_moment}
	\end{equation}
	and
	\begin{equation}
		\mathbb{E}\left(Y^3\right) = C_3 \delta^3 + C_4 \left(\sigma^2 + \delta^2\right) \delta,
		\label{eq:ST_third_moment}
	\end{equation}
	respectively. Here $C_1 = M_{2} -  M^2_{1} \left(2 / \pi\right), C_2 = M_2, C_3 = 2M_{1}^3(2/\pi)^{3/2} + 2 M_{3} \sqrt{2/\pi} - 3 M_1 M_2 \sqrt{2/\pi},$ and $C_4 = 3\left(M_{3} - M_{1}M_2\right) \sqrt{2/\pi}$. After simple algebra, we have that \eqref{eq:ST_second_moment} and \eqref{eq:ST_third_moment} imply 
	\begin{equation}
		\delta^3 + \frac{C_4 \mathbb{E}\left(Y^2\right)}{C_2 C_3 - C_1 C_4} \delta + \frac{- C_2}{C_2 C_3 - C_1 C_4} \mathbb{E}\left(Y^3\right) = 0.
		\label{eq:delta_equation_identifiability}
	\end{equation}
	Under the assumption (A4), via a computer program, we can easily verify that the ratio $\frac{C_4}{C_2 C_3 - C_1 C_4}$ is positive for any $\nu = 4, \dots, 100$. Then, we can apply the Cardano’s formula to derive the unique real root of \eqref{eq:delta_equation_identifiability}:
	$$
	\delta=\sqrt[3]{-\frac{q}{2}+\sqrt{\frac{q^2}{4}+\frac{p^3}{27}}}+\sqrt[3]{-\frac{q}{2}-\sqrt{\frac{q^2}{4}+\frac{p^3}{27}}},
	$$
	where $p = \frac{C_4}{C_2 C_3 - C_1 C_4} \mathbb{E}\left(Y^2\right),$ and $q = \frac{- C_2}{C_2 C_3 - C_1 C_4} \mathbb{E}\left(Y^3\right)$. 
	
	With the assumption, $\mathcal{M}_{1} = \mathcal{M}_{2}$, it is obvious that $\mathbb{E}\left(Y_{1}^2\right)=\mathbb{E}\left(Y_{2}^2\right)$ and $\mathbb{E}\left(Y_{1}^3\right)=\mathbb{E}\left(Y_{2}^3\right)$. With the identifiability of $\nu$ in the first step, we have that $\delta$ is identifiable.
	
	We are done.
	
\end{proof}

\begin{lemma}\label{lmm_identifiable}
	Let
	\begin{equation}
		\mathbf{Y} = \boldsymbol{\theta} + \boldsymbol{1}_{n} b + \boldsymbol{\epsilon},
		\label{eq:lmm_identifiability}
	\end{equation}
	where 
	\begin{equation}
		\left(\begin{array}{c}
			b \\
			\boldsymbol{\epsilon}
		\end{array}\right) \stackrel{\text { ind }}{\sim} \mathrm{ST}_{n + 1}\left[\left(\begin{array}{c}
			h(\nu) \delta \\
			\mathbf{0}_{n \times 1}
		\end{array}\right),\left(\begin{array}{cc}
			d^{2} & \mathbf{0}_{1 \times n} \\
			\mathbf{0}_{n \times 1} & \sigma^2 \mathbf{I}_{n}
		\end{array}\right),\left(\begin{array}{c}
			\delta \\
			\mathbf{0}_{n \times 1}
		\end{array}\right), \nu\right],
		\label{eq:lmm_distributional_assumption}
	\end{equation}
	$n > 1$,
	and
	$$h(\nu) = -\sqrt{\nu / \pi} \Gamma \left(\frac{\nu - 1}{2}\right) / \Gamma\left(\frac{\nu}{2}\right).$$
	Equivalently,
	$$
	\mathbf{Y} \sim \operatorname{ST}_{n}\left(\boldsymbol{\theta} + h\left(\nu\right)\delta \mathbf{1}_{n\times 1}, d^2 \mathbf{1}_{n\times 1} \mathbf{1}_{1\times n} + \sigma^2 \mathbf{I}_n, \delta \mathbf{1}_{n\times 1}, \nu \right).
	$$
	If the assumption (A4) holds, then all parameters are identifiable. 
\end{lemma}

\begin{proof}
	First, we want to show that $\boldsymbol{\theta}$ is identifiable. 
	
	From (7) of \citet{schumacher2021canonical}, with the assumption (A4), the first-order moment $\mathbf{Y}$ exist, and,
	$$
	\mathbb{E}\left(\mathbf{Y}\right)=\mathbb{E}\left(\boldsymbol{\theta} + \boldsymbol{1}_{n} b + \boldsymbol{\epsilon}\right) = \boldsymbol{\theta}.
	$$
	
	Let $\mathcal{M}_{1}$ and $\mathcal{M}_{2}$ represent models in \eqref{eq:lmm_identifiability} and \eqref{eq:lmm_distributional_assumption} with the parameterizations $\left[\boldsymbol{\theta}_{1},\sigma_{1},d_{1},\delta_{1},\nu_{1}\right]$ and $\left[\boldsymbol{\theta}_{2},\sigma_{2},d_{2},\delta_{2},\nu_{2}\right]$, respectively. Suppose that $\mathcal{M}_{1} = \mathcal{M}_{2}$, we have that
	$$\boldsymbol{\theta}_{1} = \boldsymbol{\theta}_{2},$$
	regardless the values of $\sigma_1,d_1,\delta_1,\nu_1,\sigma_2,d_2,\delta_2$ and $\nu_2$.
	That is, $\boldsymbol{\theta}$ is identifiable.
	
	Second, we want to show that $\nu$ and $\delta$ are identifiable. 
	
	Let $Y_1$ denote for the first element of $\mathbf{Y}$. Similarly, let $\theta_{1}$ denote for the first element of $\boldsymbol{\theta}$. The ST distribution is closed under a linear transformation (Proposition 5 of \citet{schumacher2021canonical}), such that the distribution of $Y_1 - \theta_1$ is the one in \eqref{eq:identifiability_nu_delta}. With the identifiability of $\boldsymbol{\theta}$ in the first step, we can apply Lemma \ref{skewness_identifiable} to prove that $\nu$ and $\delta$ are identifiable.
	
	Finally, we want to show that $d^2$ and $\sigma^2$ are identifiable.
	
	From (7) of \citet{schumacher2021canonical},
	$$
	\operatorname{Var} \left(\mathbf{Y}\right) = \frac{\nu}{\nu - 2} \left(d^2 \mathbf{1}_{n\times 1} \mathbf{1}_{1\times n} + \sigma^2 \mathbf{I}_n + \left(1 - \frac{2}{\pi}\right)\delta^2 \mathbf{1}_{n\times 1} \mathbf{1}_{1\times n} \right) + a\left(\nu\right) \delta^2 \mathbf{1}_{n\times 1} \mathbf{1}_{1\times n},
	$$
	where $\kappa_1=(\nu / 2)^{1 / 2} \Gamma\left(\frac{\nu-1}{2}\right) / \Gamma\left(\frac{\nu}{2}\right), a(\nu)=\frac{2}{\pi}\left(\frac{\nu}{\nu-2}-\kappa_1^2\right)$. Suppose $\mathcal{M}_{1} = \mathcal{M}_{2}$, then $\operatorname{Var}\left(\mathbf{Y}_{1}\right) = \operatorname{Var}\left(\mathbf{Y}_{2}\right)$. With the identifiability of $\nu$ and $\delta$ from the second step, $\operatorname{Var}\left(\mathbf{Y}_{1}\right) = \operatorname{Var}\left(\mathbf{Y}_{2}\right)$ implies
	$$
	d_1^2 \mathbf{1}_{n\times 1} \mathbf{1}_{1\times n} + \sigma_1^2 \mathbf{I}_n = d_2^2 \mathbf{1}_{n\times 1} \mathbf{1}_{1\times n} + \sigma_2^2 \mathbf{I}_n.
	$$
	That is $d_1 = d_2$ and $\sigma_1 = \sigma_2$. Equivalently, $d^2$ and $\sigma^2$ are identifiable.
	
	We are done.
\end{proof}

\begin{lemma}\label{sim_identifiable}
	Let 
	\begin{equation}
		\mathbf{Y} = g\left(\mathbf{X}\boldsymbol{\beta} \right) + \boldsymbol{1}_{n} b + \boldsymbol{\epsilon},
		\label{eq:single_index_identifiability}
	\end{equation}
	with $n > 1$, and
	$$
	g\left(\mathbf{X}\boldsymbol{\beta}\right) = 
	\begin{pmatrix}
		g^{\star}\left(\mathbf{X}^{(1)}\boldsymbol{\beta}\right) \\
		\vdots \\
		g^{\star}\left(\mathbf{X}^{(n)}\boldsymbol{\beta}\right)
	\end{pmatrix},
	$$
	where $\mathbf{X}^{(1)}$ represents the first row of the $\mathbf{X}$. The distributional assumption remains the same as in \eqref{eq:lmm_distributional_assumption}.
	
	If the assumptions (A1), (A2), (A3) and (A4) hold, then $g(\cdot)$, $g^{\star}(\cdot)$, $\boldsymbol{\beta}, \nu, \delta, d^2$ and $\sigma^2$ are identifiable.
\end{lemma}

\begin{proof}
	While a similar proof is given by \citet{lin2007identifiability}, we provide the following for completeness.
	
	Let $\mathcal{M}_{1}$ and $\mathcal{M}_{2}$ represent models in \eqref{eq:single_index_identifiability}  with parameterization $\left[\boldsymbol{\beta}_1,g^{\star}_{1}(\cdot),\sigma_{1},d_{1},\delta_{1},\nu_{1}\right]$ and $\left[\boldsymbol{\beta}_2,g^{\star}_{2}(\cdot),\sigma_{2},d_{2},\delta_{2},\nu_{2}\right]$, respectively. Then, with the assumption (A4), which ensures the existence of first-order moment of $\mathbf{Y}$, by Lemma \ref{lmm_identifiable}, $\mathcal{M}_{1} = \mathcal{M}_{2}$ implies that
	$$\mathbb{E}\left(\mathbf{Y}_{1}\right) = g_{1}\left(\mathbf{X}\boldsymbol{\beta}_{1} \right) = \mathbb{E}\left(\mathbf{Y}_{2}\right) = g_{2}\left(\mathbf{X}\boldsymbol{\beta}_{2} \right),$$
	and
	$$
	m\left(\mathbf{X}^{(j)}\right) = g_{1}^{\star}\left(\mathbf{X}^{(j)}\boldsymbol{\beta}_{1}\right) = g_{2}^{\star}\left(\mathbf{X}^{(j)}\boldsymbol{\beta}_{2}\right) \quad \text{for } j = 1, \dots, n, 
	$$
	regardless the values of $\sigma_1, d_1, \delta_1, \nu_1,\sigma_2, d_2, \delta_2$ and $\nu_2$.
	To simplify notation, let $X$ represent a transpose $\mathbf{X}^{(j)}$ for a given $j = 1,\dots,n.$
	
	We want to show that, under assumptions (A1), (A2) and (A3), if
	$$
	m\left(X\right) = g_{1}^{\star}\left(\boldsymbol{\beta}_{1}^{\top} X\right) = g_{2}^{\star}\left(\boldsymbol{\beta}_{2}^{\top} X\right), \quad \text{for all } X \in S
	$$
	then $\boldsymbol{\beta}_{1} = \boldsymbol{\beta}_{2}$. Here $S$ represents the support of $m(\cdot)$.
	
	Suppose $\boldsymbol{\beta}_{1} \neq \boldsymbol{\beta}_{2}$. Under the assumption (A1), there exists a sphere $B = B(X_{0},r) \subset S$ for some $X_{0}$ such that $X_{0} + t\boldsymbol{\beta}_{1} \in S$ , $X_{0} + t\boldsymbol{\beta}_{2} \in S$ for all $t\in (-r,r)$. By the assumption (A3), $\boldsymbol{\beta}^{\top}_{1}\boldsymbol{\beta}_{1} = \boldsymbol{\beta}^{\top}_{2}\boldsymbol{\beta}_{2} =1$, we have that
	$$
	g_{1}^{\star}\left(\boldsymbol{\beta}_{1}^{\top} X_{0} + t\right) = g_{1}^{\star}\left(\boldsymbol{\beta}_{1}^{\top} \left(X_{0} + t\boldsymbol{\beta}_{1}\right)\right) = g_{2}^{\star}\left(\boldsymbol{\beta}_{2}^{\top} \left(X_{0} + t\boldsymbol{\beta}_{1}\right)\right) = g_{2}^{\star}\left(\boldsymbol{\beta}_{2}^{\top} X_{0} + t\boldsymbol{\beta}_{2}^{\top}\boldsymbol{\beta}_{1}\right),
	$$
	$$
	g_{2}^{\star}\left(\boldsymbol{\beta}_{2}^{\top} X_{0} + t\right) = g_{2}^{\star}\left(\boldsymbol{\beta}_{2}^{\top} \left(X_{0} + t\boldsymbol{\beta}_{2}\right)\right) = g_{1}^{\star}\left(\boldsymbol{\beta}_{1}^{\top} \left(X_{0} + t\boldsymbol{\beta}_{2}\right)\right) = g_{1}^{\star}\left(\boldsymbol{\beta}_{1}^{\top} X_{0} + t\boldsymbol{\beta}_{2}^{\top}\boldsymbol{\beta}_{1}\right).
	$$
	By the Cauchy-Schwarz inequality, $|\boldsymbol{\beta}_{1}^{\top} \boldsymbol{\beta}_{2}| < 1$. By the continuity assumption from (A2),
	$$
	\begin{aligned}
		g_{1}^{\star}\left(\boldsymbol{\beta}_{1}^{\top} X_{0} + t\right) &= g_{2}^{\star}\left(\boldsymbol{\beta}_{2}^{\top} X_{0} + t\boldsymbol{\beta}_{2}^{\top}\boldsymbol{\beta}_{1}\right) \\
		&= g_{1}^{\star}\left(\boldsymbol{\beta}_{1}^{\top} X_{0} + t\left(\boldsymbol{\beta}_{2}^{\top}\boldsymbol{\beta}_{1}\right)^2\right)\\
		&= \cdots \\
		&= g_{1}^{\star}\left(\boldsymbol{\beta}_{1}^{\top} X_{0} + t\left(\boldsymbol{\beta}_{2}^{\top}\boldsymbol{\beta}_{1}\right)^{2n}\right) \\
		&= \cdots \\
		&= g_{1}^{\star}\left(\boldsymbol{\beta}_{1}^{\top} X_{0}\right), \quad \text{for all } t \in (-r,r).
	\end{aligned}
	$$
	Note that $g_{1}^{\star}\left(\boldsymbol{\beta}_{1}^{\top} X_{0} + t\right) = g_{1}^{\star}\left(\boldsymbol{\beta}_{1}^{\top} X_{0}\right) \text{ for all } t \in (-r,r)$, contradicts with the monotonic increasing assumption of  $g_{1}(\cdot)$. Therefore $\boldsymbol{\beta}_{1} = \boldsymbol{\beta}_{2}$ must hold.
	
	With $\boldsymbol{\beta}_{1} = \boldsymbol{\beta}_{2}$, it is obvious that $g_{1}^{\star}(\cdot) = g_{2}^{\star}(\cdot)$ and $g_{1}(\cdot) = g_{2}(\cdot)$.
	
	Last, with the identifiability of $\boldsymbol{\beta}$ and $g^{\star}(\cdot)$, the identifiability of $\nu, \delta, d^2$ and $\sigma^2$ is implied by Lemma \ref{lmm_identifiable}.
	
	We are done.
\end{proof}

Finally, with Lemma \ref{lmm_identifiable} and Lemma \ref{sim_identifiable}, we can prove Theorem 1.

\begin{proof}
	We want to prove identifiability of the following model:
	\begin{equation}
		\mathbf{Y}=\left(\begin{array}{c}
			\mathbf{Y}^P \\
			\mathbf{Y}^C
		\end{array}\right)=\left(\begin{array}{c}
			g\left(\mathbf{X} \boldsymbol{\beta}\right) \\
			a \times g\left(\mathbf{X} \boldsymbol{\beta}\right)
		\end{array}\right)+\left(\begin{array}{c}
			\boldsymbol{1}_{n} \\
			\boldsymbol{1}_{n}
		\end{array}\right) b+\left(\begin{array}{c}
			\boldsymbol{\epsilon}^P \\
			\boldsymbol{\epsilon}^C
		\end{array}\right),
		\label{eq:ST-GP_model_appendix}
	\end{equation}
	where the slope parameter $a \in (-\infty,\infty)$. The $g(\cdot)$ remains the same as defined in Lemma \ref{sim_identifiable}. The distributional assumption for the random effects and errors is expressed as follows:
	\begin{equation}
		\left(\begin{array}{c}
			b \\
			\boldsymbol{\epsilon}
		\end{array}\right) \sim ST_{2n+1}\left[\left(\begin{array}{c}
			h(\nu) \delta \\
			\boldsymbol{0}_{2n \times 1}
		\end{array}\right),\left(\begin{array}{cc}
			d^2 & \boldsymbol{0}_{1 \times 2n} \\
			\boldsymbol{0}_{2n \times 1} & \sigma^2 \mathbf{I}_{2n \times 2n}
		\end{array}\right),\left(\begin{array}{c}
			\delta \\
			\boldsymbol{0}_{2n \times 1}
		\end{array}\right), \nu\right],
		\label{eq:single_index_model_distributional_assumption_appendix}
	\end{equation}
	where $n > 1$, and
	$$\boldsymbol{\epsilon} = 
	\left(\begin{array}{l}
		\boldsymbol{\epsilon}^P \\
		\boldsymbol{\epsilon}^C
	\end{array}\right).$$
	
	First, we want to show that $\boldsymbol{\beta}$, $g(\cdot)$ and $g^{\star}$ are identifiable regardless the values of other parameters. Let $\mathcal{M}_{1}$ and $\mathcal{M}_{2}$ represent models in \eqref{eq:ST-GP_model_appendix} and \eqref{eq:single_index_model_distributional_assumption_appendix} with parameterization $\left[a_{1},g_{1},g^{\star}_{1},\boldsymbol{\beta}_{1},\sigma_{1},d_{1},\delta_{1},\nu_{1}\right]$ and $\left[a_{2},g_{2},g^{\star}_{2},\boldsymbol{\beta}_{2},\sigma_{2},d_{2},\delta_{2},\nu_{2}\right]$, respectively. Then, $\mathcal{M}_{1} = \mathcal{M}_{2}$ implies that $\mathbf{Y}_{1}$ is equivalent to $\mathbf{Y}_{2}$ in distribution, denoted as $\mathbf{Y}_{1} \stackrel{d}{=} \mathbf{Y}_{2}$. By Proposition 5 of \citet{schumacher2021canonical}, this implies that $\mathbf{Y}^{P}_{1} \stackrel{d}{=} \mathbf{Y}^{P}_{2}$. Then, under assumptions (A1), (A2), (A3) and (A4), by Lemma \ref{sim_identifiable}, $\boldsymbol{\beta},g(\cdot)$ and $g^{\star}(\cdot)$ are identifiable.
	
	Second, we want to show that $a$ is identifiable. Again, from $\mathbf{Y}_{1} \stackrel{d}{=} \mathbf{Y}_{2}$ and Proposition 5 of \citet{schumacher2021canonical}, we have that $\mathbf{Y}^{C}_{1} \stackrel{d}{=} \mathbf{Y}^{C}_{2}$. Note that $\mathbb{E}\left(\mathbf{Y}^{C}\right) = a \times g\left(\mathbf{X} \boldsymbol{\beta}\right)$. With the identifiability of $g(\cdot)$ and $\boldsymbol{\beta}$ from the first step, we have that $a$ is identifiable.
	
	Last, with the identifiability of $a, \boldsymbol{\beta}$ and $g^{\star}(\cdot)$, the identifiability of $\nu, \delta, d^2$ and $\sigma^2$ is implied by Lemma \ref{lmm_identifiable}.
	
	We are done.
\end{proof}

\section{The PBS and WFPBB Algorithms} \label{supp:PBS_and_WFPBB}

\begin{algorithm}
	\caption{\label{alg:WFPBB}WFPBB algorithm.}
	\begin{algorithmic}[1]
		\Procedure{WFPBB$\left(\mathbf{Y},\boldsymbol{w}, N,n\right)$.}{}
		\State $l_{i} \gets 0 \quad \forall i = 1,\dots,n$;
		\For{$k = 1: (N-n)$}
		\State Letting $N^{\star} = (N-n)/n$, draws $Y^{\star}_{k} = Y_{i}$ with probability
		$$
		\frac{w_i-1+l_i N^*}{N-n+(k-1) \times N^*},
		$$
		\If{$Y^{\star}_{k} = Y_{i}$}
		\State $l_{i} \gets l_{i} + 1$;
		\EndIf
		\EndFor
		\State Stack $\left(Y_{1},Y_{2},\dots,Y_{n}\right)$ and $\left(Y^{\star}_{1},Y^{\star}_{2},\dots,Y^{\star}_{N-n}\right)$ to form a pseudo population;
		\State Randomly draw a sample of size $n$ from the pseudo population;
		\EndProcedure.
	\end{algorithmic}
\end{algorithm}

\begin{algorithm}
	\caption{\label{alg:PRB}Parallel MCMC with PRS.}
	\begin{algorithmic}[1]
		\Procedure{MCMC-PRS$\left(\mathbf{Y},\boldsymbol{w},M,J\right)$.}{}
		\For{j = 1 : $J$} \Comment{This loop can be done in parallel}
		\State Draw $\mathbf{Z}^{[j]}$ from $p\left(\mathbf{Z} \mid \mathbf{Y}, \boldsymbol{w}\right)$;
		\For{i = 1: $M$}
		\State Draw $\boldsymbol{\theta}^{[i]}$ from $p\left(\boldsymbol{\theta} \mid \mathbf{Z}^{[j]} \right)$.
		\EndFor
		\EndFor
		\EndProcedure.
	\end{algorithmic}
\end{algorithm}

In Algorithm \ref{alg:WFPBB}, $N$ stands for the population size.

\section{Complete Index Calculation Formula} \label{supp:Complete Index Calculation Formula}

{\small
\begin{equation}
\begin{aligned}
    \hat{U} = &- \mathbbm{1}\left(\text{Female}\right) \times \frac{1-0.508}{0.5} \times 0.102 
    - \mathbbm{1}\left(\text{Male}\right) \times \frac{0-0.508}{0.5} \times 0.102 \\
    & - \mathbbm{1}\left(\text{Diabetes:no}\right) \times \frac{1-0.886}{0.317} \times 0.074 
    - \mathbbm{1}\left(\text{Diabetes:yes}\right) \times \frac{0-0.886}{0.317} \times 0.074 \\
    & - \mathbbm{1}\left(\text{Upperjaw:no}\right) \times \frac{1-0.507}{0.5} \times 0.018 
    - \mathbbm{1}\left(\text{Upperjaw:yes}\right) \times \frac{0-0.507}{0.5} \times 0.018 \\
    & - \mathbbm{1}\left(\text{Interproximal Area:no}\right) \times \frac{1-0.335}{0.472} \times 0.469 
    - \mathbbm{1}\left(\text{Interproximal Area:yes}\right) \times \frac{0-0.335}{0.472} \times 0.469 \\
    & - \mathbbm{1}\left(\text{Molar:no}\right) \times \frac{1-0.753}{0.431} \times 0.553 
    - \mathbbm{1}\left(\text{Molar:yes}\right) \times \frac{0-0.753}{0.431} \times 0.553 \\
    & + \frac{\text{Age} - 50.575}{13.966} \times 0.006  + \frac{\text{Ratio of Family Income to Poverty} - 2.824}{1.662} \times (-0.023) \\
    & + \frac{\text{BMI} - 29.374}{6.654} \times 0.002 
     + \frac{\text{HDL Cholesterol} - 53.090}{16.105} \times 0.022 \\
    & + \frac{\text{Total Cholesterol} - 198.160}{41.846} \times 0.007 
     + \frac{\text{Glycohemoglobin Percentage} - 5.745}{1.019} \times (-0.018) \\
    & + \frac{\text{Blood Lead} - 1.520}{1.798} \times 0.043 
     + \frac{\text{Healthy Eating Index} - 52.883}{13.928} \times (-0.003) \\
    & + \mathbbm{1}\left(\text{Binge Drinking:no}\right) \times \frac{1-0.252}{0.434} \times 0.011 
     + \mathbbm{1}\left(\text{Binge Drinking:yes}\right) \times \frac{0-0.252}{0.434} \times 0.011 \\
    & + \mathbbm{1}\left(\text{Health Insurance:no}\right) \times \frac{1-0.216}{0.411} \times 0.023 
     + \mathbbm{1}\left(\text{Health Insurance:yes}\right) \times \frac{0-0.216}{0.411} \times 0.023 \\
    & - \mathbbm{1}\left(\text{Tobacco Intake:no}\right) \times \frac{1-0.807}{0.395} \times 0.034 
     - \mathbbm{1}\left(\text{Tobacco Intake:yes}\right) \times \frac{0-0.807}{0.395} \times 0.034 \\
    & - \mathbbm{1}\left(\text{Hypertension:no}\right) \times \frac{1-0.651}{0.477} \times 0.006 
     - \mathbbm{1}\left(\text{Hypertension:yes}\right) \times \frac{0-0.651}{0.477} \times 0.006 \\
    & - \mathbbm{1}\left(\text{Race:white}\right) \times \frac{1-0.469}{0.499} \times 0.070 
     - \mathbbm{1}\left(\text{Race:not white}\right) \times \frac{0-0.469}{0.499} \times 0.070 \\
    & + \mathbbm{1}\left(\text{Race:black}\right) \times \frac{1-0.184}{0.387} \times 0.026 
     + \mathbbm{1}\left(\text{Race:not black}\right) \times \frac{0-0.184}{0.387} \times 0.026 \\
    & + \mathbbm{1}\left(\text{Race:Hispanic}\right) \times \frac{1-0.237}{0.425} \times 0.040 
     + \mathbbm{1}\left(\text{Race:not Hispanic}\right) \times \frac{0-0.237}{0.425} \times 0.040 \\
    & - \mathbbm{1}\left(\text{Education:more than high school}\right) \times \frac{1-0.607}{0.488} \times 0.032 
    - \mathbbm{1}\left(\text{Education:high school or less}\right) \times \frac{0-0.607}{0.488} \times 0.032 \\
    & + \mathbbm{1}\left(\text{Marital Status:married/living with partner}\right) \times \frac{1-0.671}{0.470} \times 0.014 
    + \mathbbm{1}\left(\text{Marital Status:other}\right) \times \frac{0-0.671}{0.470} \times 0.014.
\end{aligned}
\label{eq:real_data_analysis_single_index_formula}
\end{equation}
}

\section{Extra Tables and Figures} \label{supp:tables_and_figures}

\begin{figure}
\centering
\begin{minipage}{0.49\textwidth}
\includegraphics[width = \textwidth]{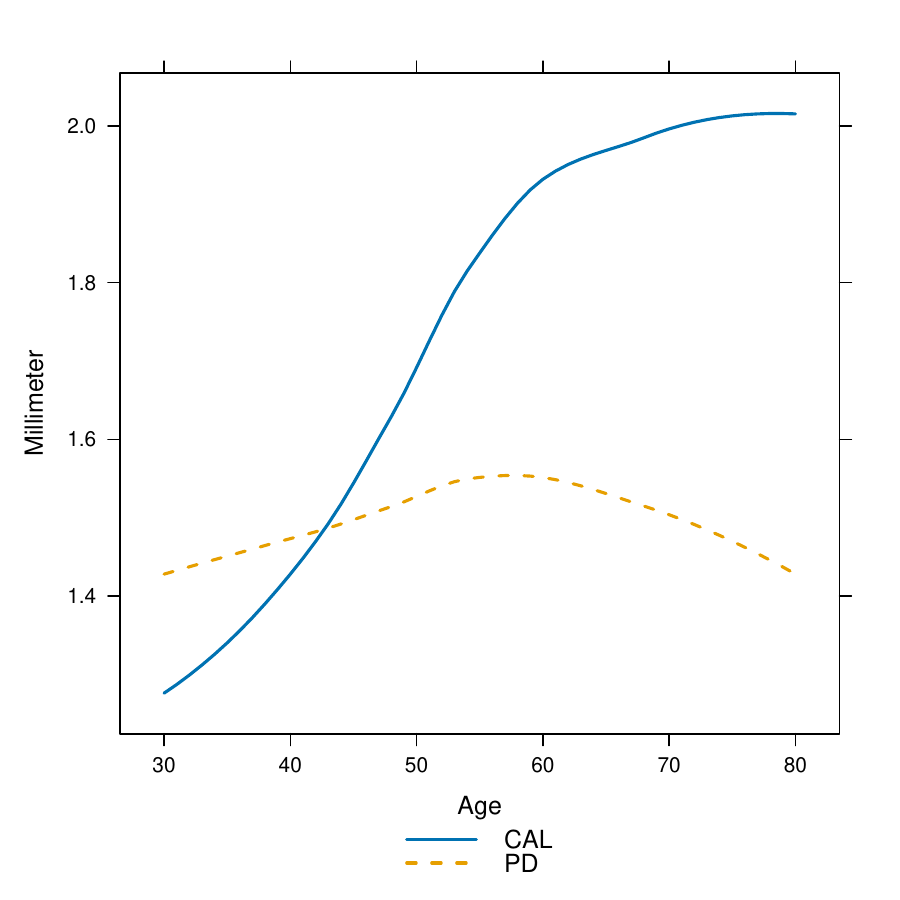}
\caption{\label{fig:real_data_analysis_LOESS_plot}NHANES data: LOESS regression model fitted to NHANES data with age as the covariate and CAL, PD as response variables, respectively.}
\end{minipage}
\begin{minipage}{0.49\textwidth}
\includegraphics[width = \textwidth]{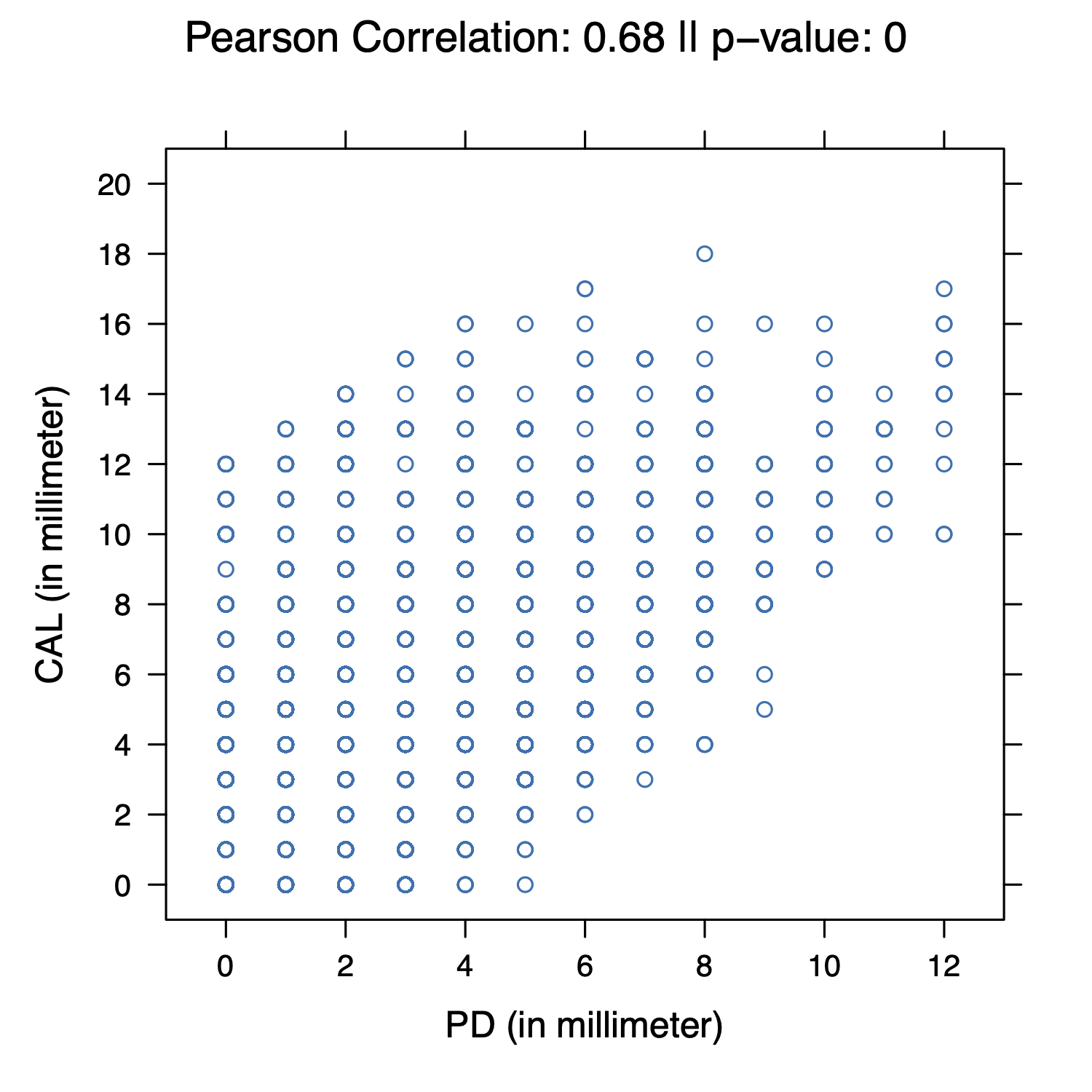}
\caption{\label{fig:real_data_analysis_CAL_PD_scatter_plot}NHANES data: Scatter plot of PD and CAL. The $p$-value in the title comes from the Pearson correlation test with the null hypothesis that the true correlation is 0.}
\end{minipage}
\end{figure}

\begin{figure}
	\centering
	\includegraphics[width = 0.98\textwidth]{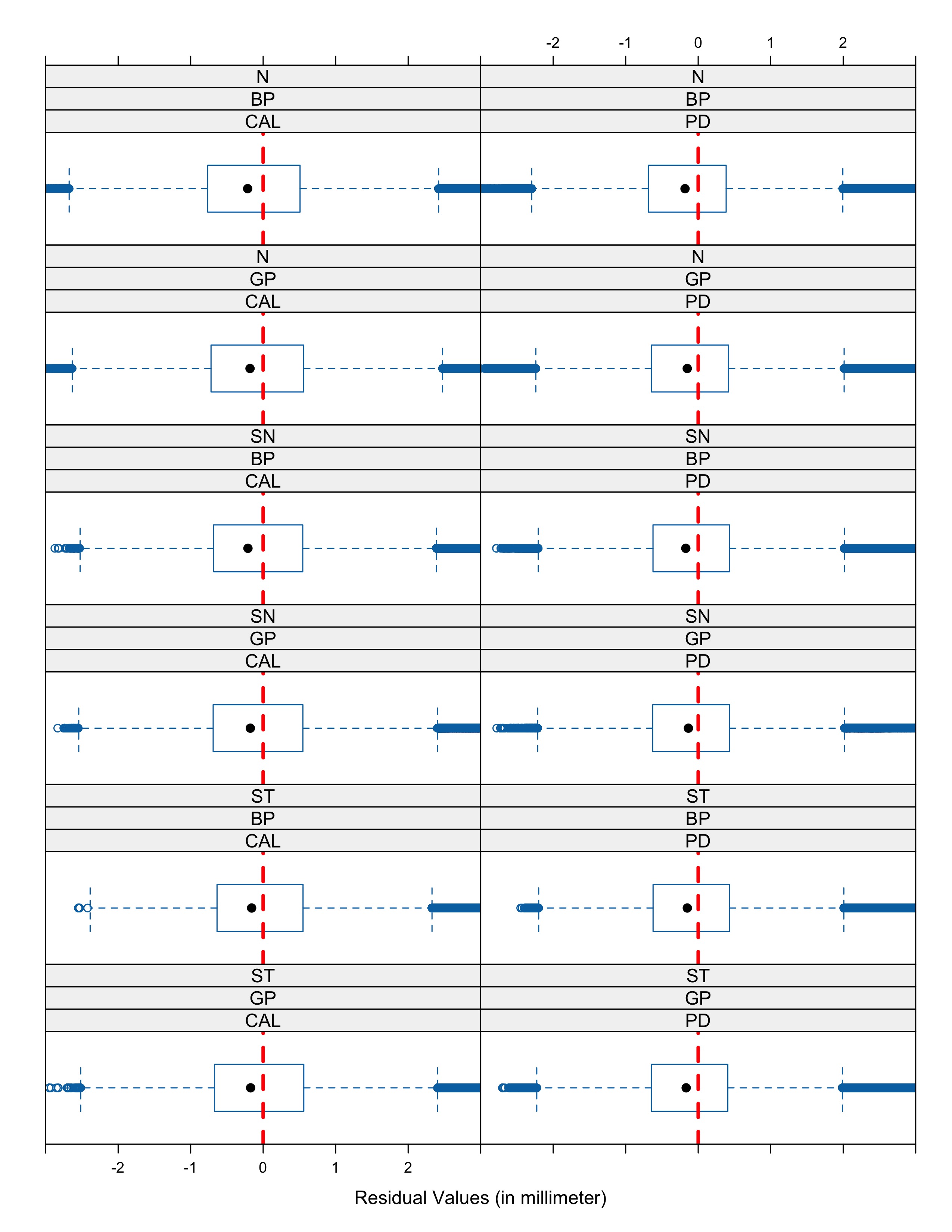}
	\caption{\label{fig:residual_six_models}NHANES data: Boxplots of residuals of PD and CAL from the ST-GP, SN-GP, N-GP, ST-BP, SN-BP and N-BP models. The red dashed lines represent the reference value (0) of residuals.}
\end{figure}

\begin{figure}
	\centering
	\includegraphics[width = \textwidth]{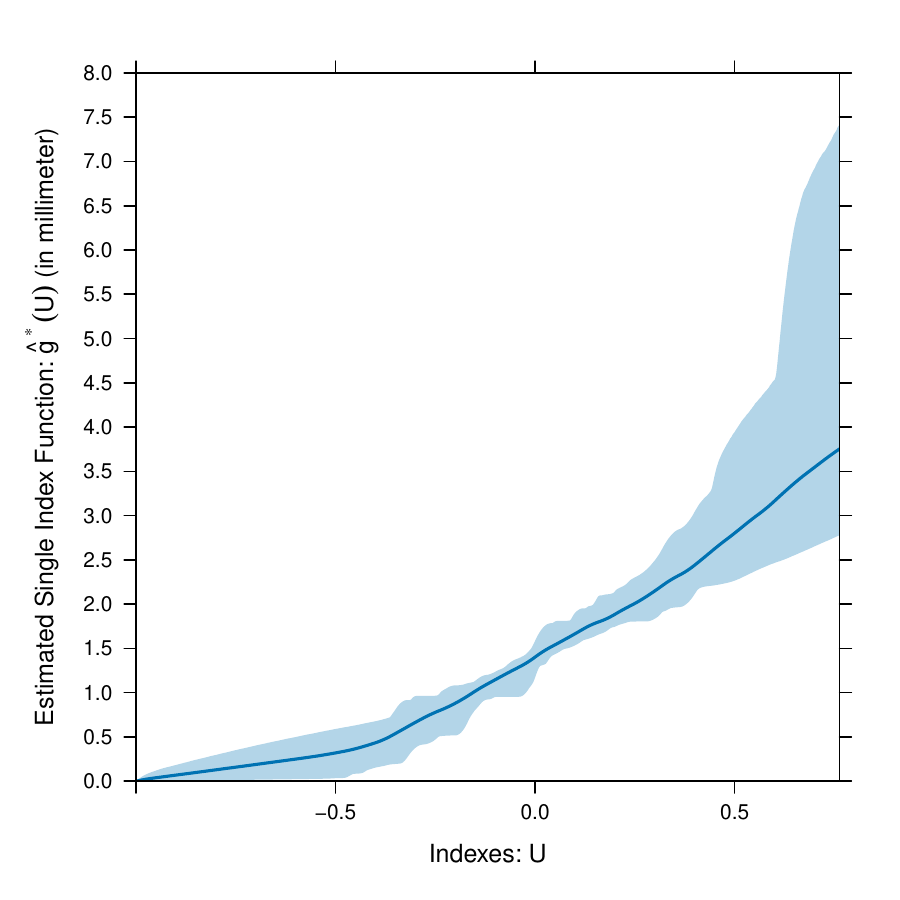}
	\caption{\label{fig:index_function}NHANES data: Estimated index function plot. The Y-axis represents the estimated $g$ function values (in millimeter). The blue transparent band depicts the $95\%$ credible interval. The solid line represents the estimated single index function.} 
\end{figure}

\begin{figure}
	\centering
	\includegraphics[width = \textwidth]{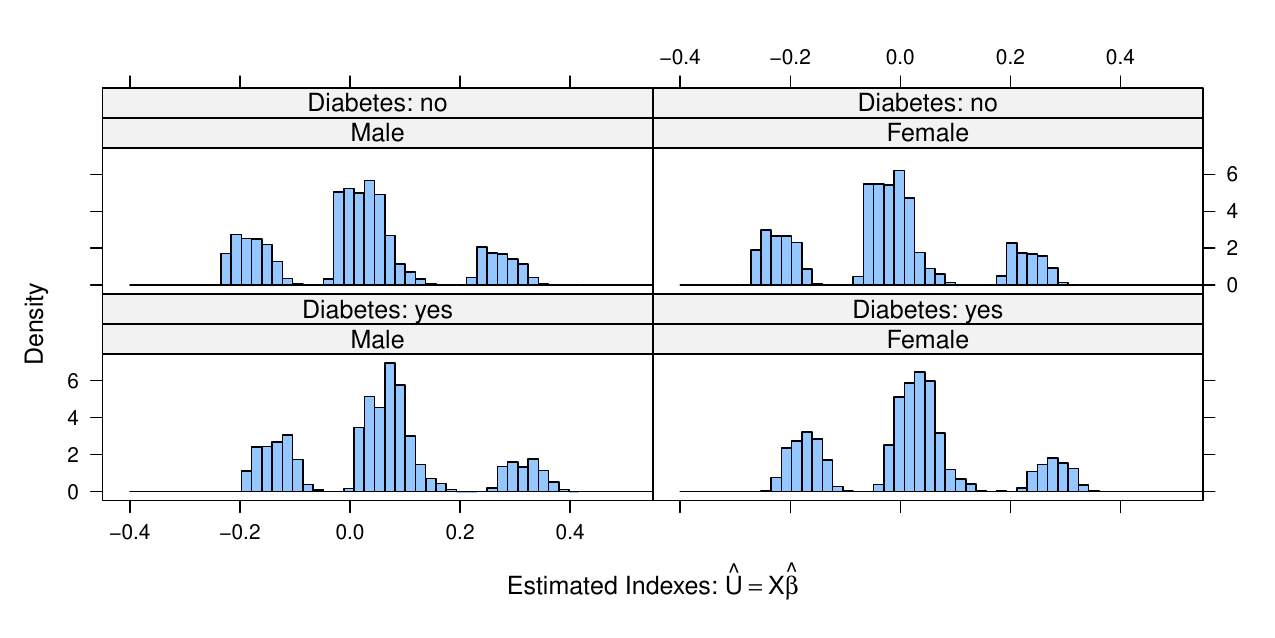}
	\caption{\label{fig:estimated indexes by gender and diabetes}NHANES data: Estimated indexes stratified by gender and diabetes status.}
\end{figure}

\begin{figure}
	\centering
	\includegraphics[width = \textwidth]{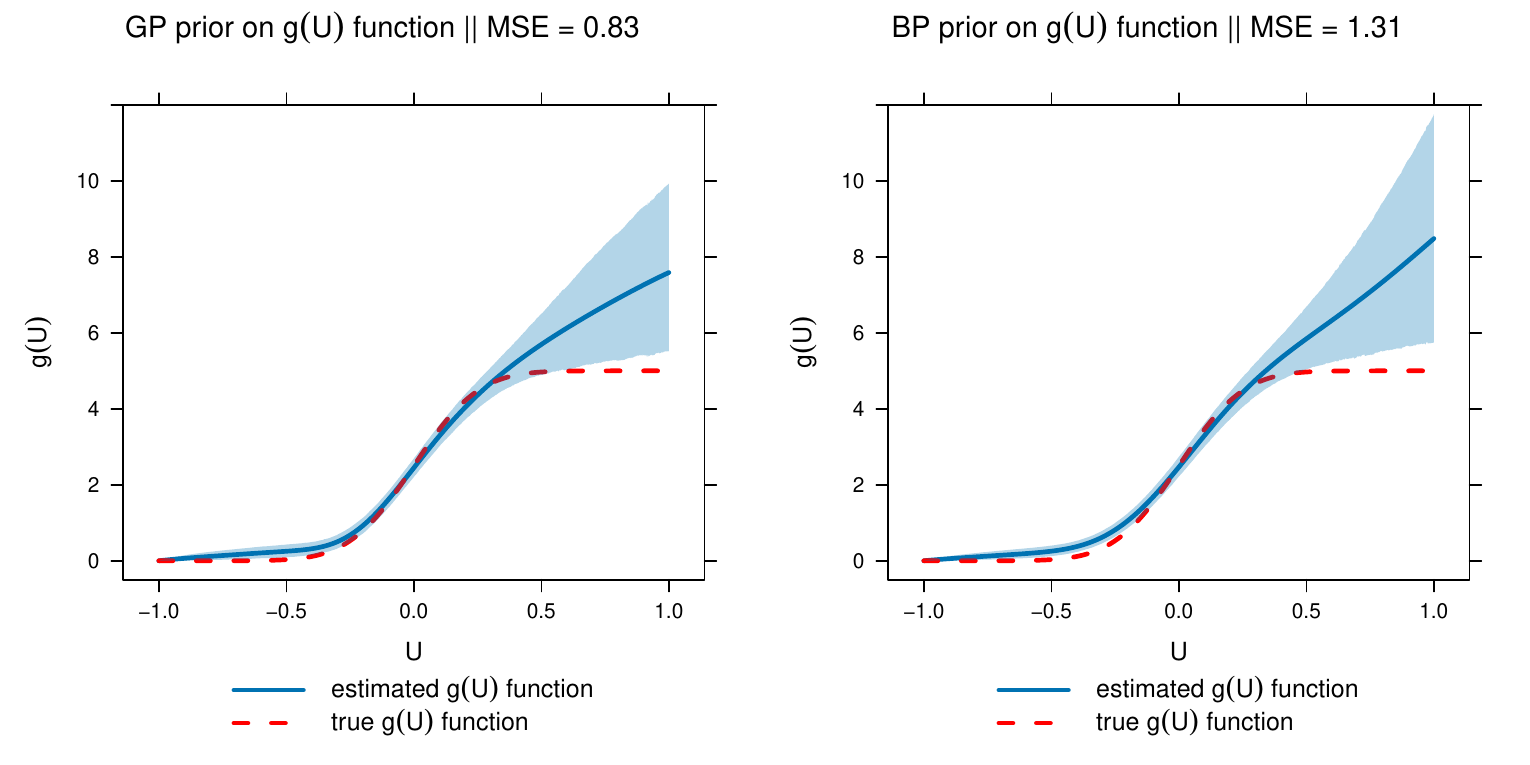}
	\caption{\label{fig:simulation_1a}Comparison between the constrained GP prior and BP prior on the index function in the first part of simulation 1. The blue solid lines and red dashed lines represent the estimated index function and true index function, respectively. Blue transparent bands depict the $95\%$ credible intervals. Green dots indicate the observed index values.}
\end{figure}

\begin{figure}
	\centering
	\includegraphics[width = 0.99\textwidth]{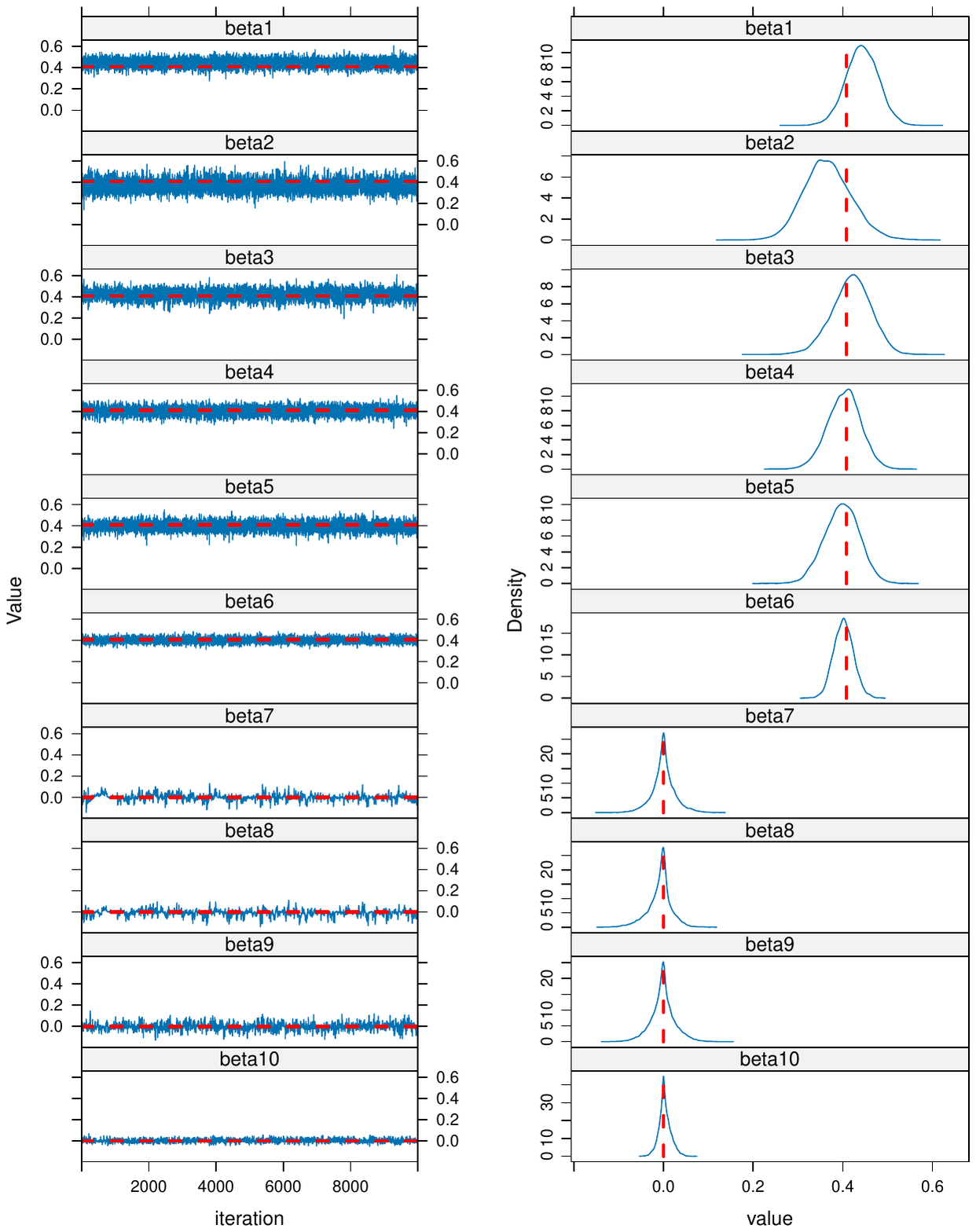}
	\caption{\label{fig:variable_selection_plot}Traceplots and density plots of the posterior distribution based on MCMC samples of $\boldsymbol{\beta}$ in the first part of simulation 1. The red dashed lines in both left and right panels represent the true values of $\boldsymbol{\beta}$.}
\end{figure}

\begin{figure}
\centering
\includegraphics[width = \textwidth]{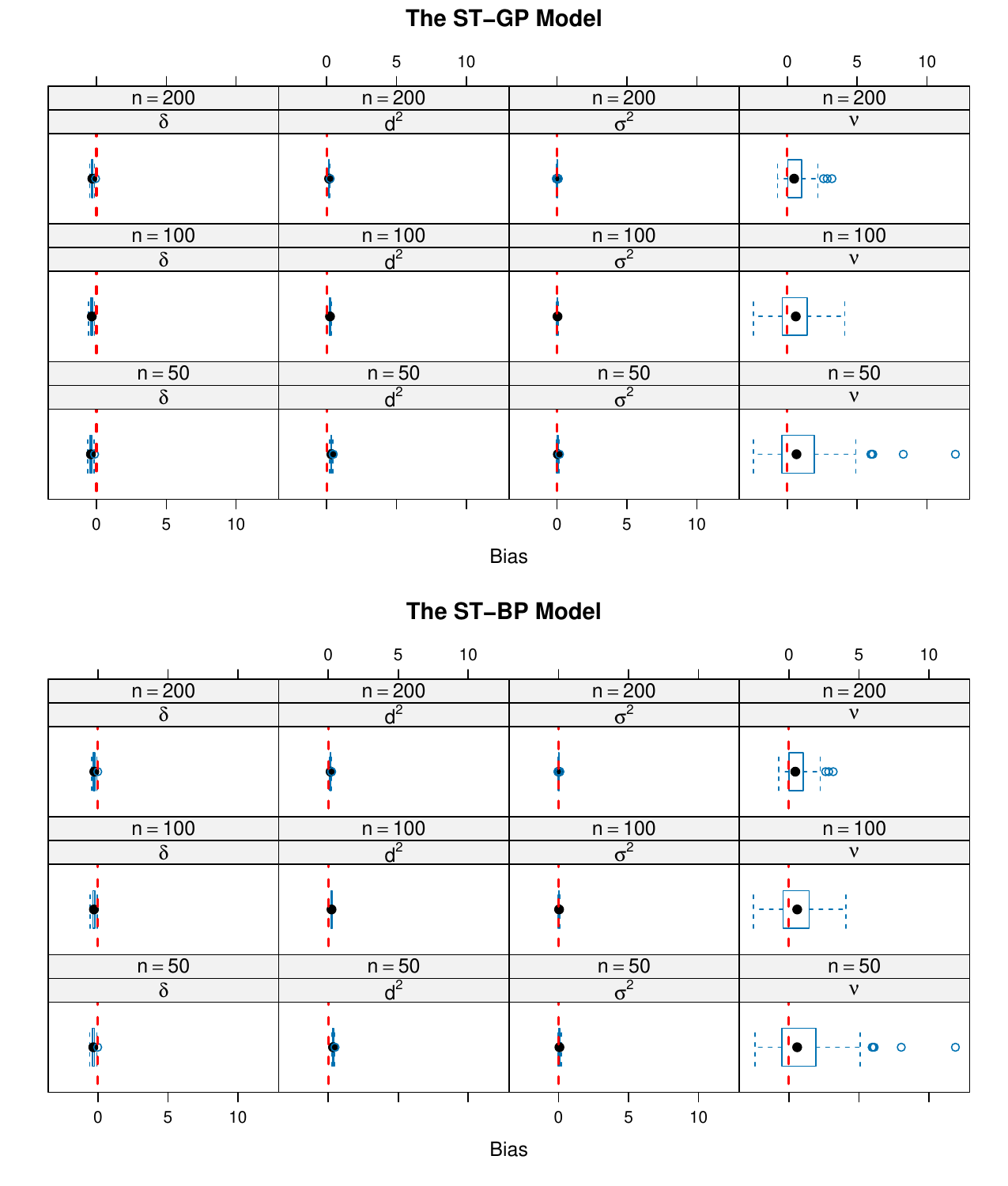}
\caption{\label{fig:boxplot of bias of other parameters} Boxplots showing the bias of all parameters, excluding those in the fixed effects, for the second part of simulation 1. Red dashed lines represent the reference value (0) of bias.}
\end{figure}

\begin{figure}
	\centering
	\includegraphics[width = \textwidth]{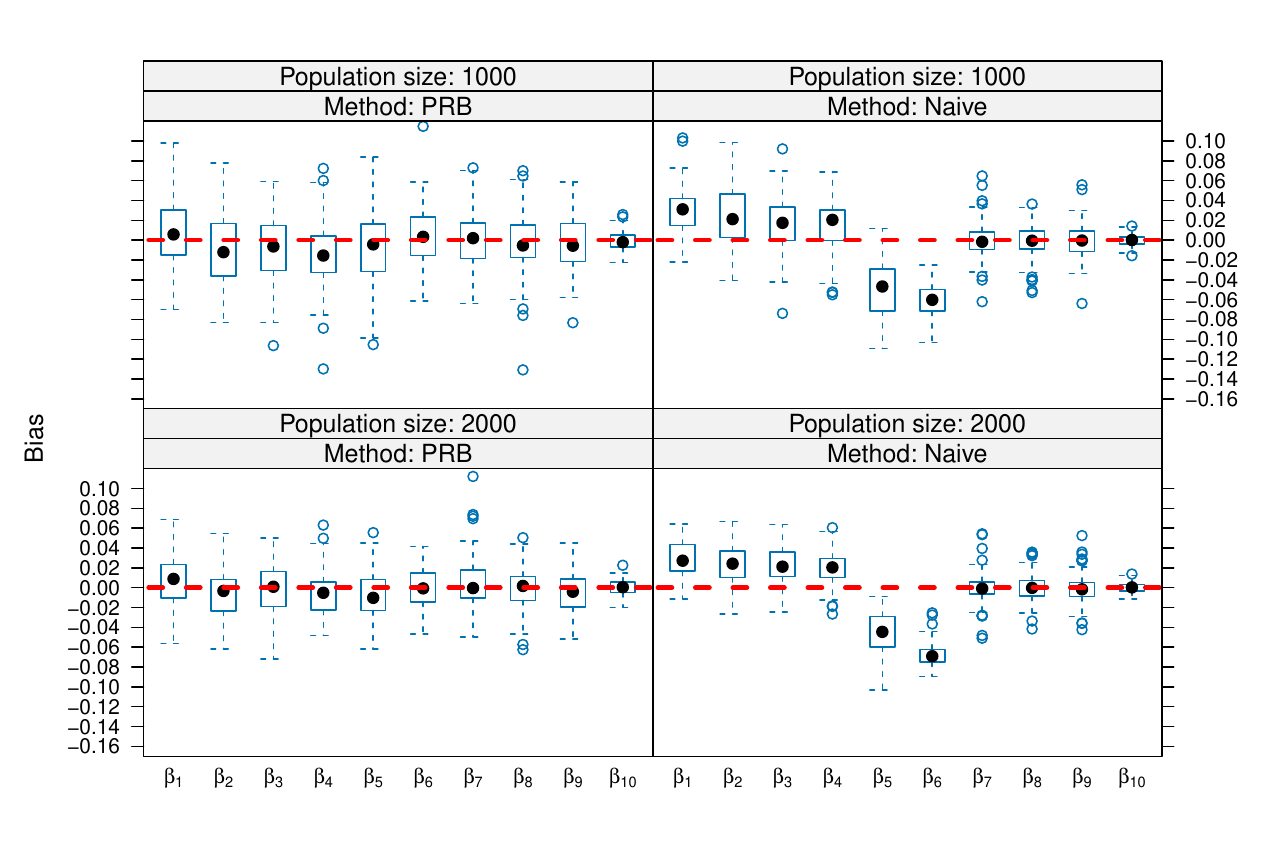}
	\caption{\label{fig:simulation2}Simulation 2: Boxplots of point estimation of $\boldsymbol{\beta}$. Red dashed lines represent the reference value (0) of bias.}
\end{figure}

\begin{figure}
	\centering
	\includegraphics[width = \textwidth]{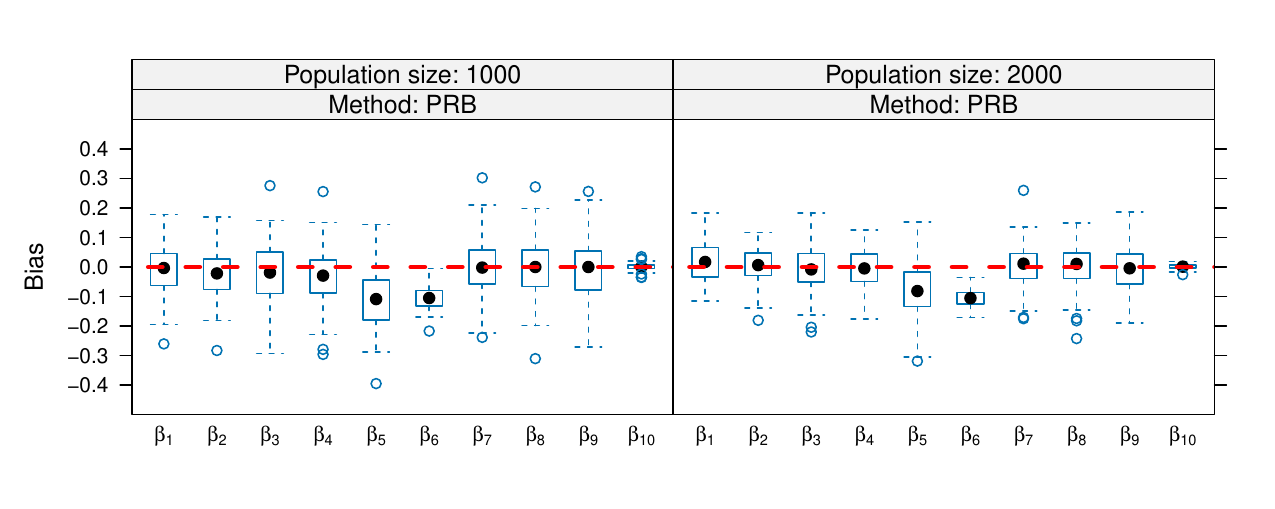}
	\caption{\label{fig:simulation3}Simulation 3: Boxplots of point estimation of $\boldsymbol{\beta}$. Red dashed lines represent the reference value (0) of bias.}
\end{figure}

\begin{table}
\caption{\label{tab:simulation1}The results from the second part of simulation 1. Except for the last row, numbers outside the parentheses represent the average bias, and numbers inside the parentheses represent 100 times the standard deviation of bias, across 100 Monte Carlo replicates. In the last row, numbers outside the parentheses represent the average of mean square errors, and numbers inside the parentheses represent the standard deviation of mean square errors, across 100 Monte Carlo replicates.}
\centering
\begin{tabular}[t]{lrrrrrr}
\toprule
\multicolumn{1}{c}{ } & \multicolumn{2}{c}{N = 50} & \multicolumn{2}{c}{N = 100} & \multicolumn{2}{c}{N = 200} \\
\cmidrule(l{3pt}r{3pt}){2-3} \cmidrule(l{3pt}r{3pt}){4-5} \cmidrule(l{3pt}r{3pt}){6-7}
  & GP50 & BP50 & GP100 & BP100 & GP200 & BP200\\
\midrule
$\alpha$ & 0.01(2.54) & 0.00(2.49) & 0.01(1.59) & 0.00(1.57) & 0.00(1.24) & 0.00(1.24)\\
$\beta_{1}$ & 0.00(3.02) & 0.00(3.08) & 0.00(2.02) & 0.01(2.01) & 0.00(1.36) & 0.00(1.39)\\
$\beta_{2}$ & 0.00(3.37) & 0.00(3.51) & 0.00(1.97) & 0.00(1.97) & -0.01(1.32) & 0.00(1.32)\\
$\beta_{3}$ & -0.02(2.91) & -0.02(2.98) & -0.01(2.12) & -0.02(2.10) & 0.00(1.37) & -0.01(1.38)\\
$\beta_{4}$ & -0.02(3.12) & -0.02(3.24) & -0.01(2.09) & -0.01(2.13) & 0.00(1.43) & -0.01(1.47)\\
\addlinespace
$\beta_{5}$ & 0.00(3.52) & -0.01(3.57) & 0.00(2.06) & 0.00(2.08) & 0.00(1.28) & 0.00(1.27)\\
$\beta_{6}$ & 0.02(1.76) & 0.02(1.90) & 0.01(1.31) & 0.01(1.32) & 0.01(0.94) & 0.01(0.93)\\
$\beta_{7}$ & 0.00(1.74) & 0.00(1.71) & 0.00(1.44) & 0.00(1.31) & 0.00(0.96) & 0.00(1.16)\\
$\beta_{8}$ & 0.00(1.78) & 0.00(1.92) & 0.00(1.33) & 0.00(1.32) & 0.00(1.11) & 0.00(1.04)\\
$\beta_{9}$ & 0.00(1.92) & 0.00(2.04) & 0.00(1.45) & 0.00(1.55) & 0.00(1.06) & 0.00(1.06)\\
\addlinespace
$\beta_{10}$ & 0.00(1.23) & 0.00(1.28) & 0.00(0.90) & 0.00(0.92) & 0.00(0.71) & 0.00(0.72)\\
$g(\cdot)$ & 0.57(0.32) & 1.14(0.70) & 0.42(0.17) & 0.81(0.18) & 0.37(0.15) & 0.80(0.19)\\
\bottomrule
\end{tabular}
\end{table}

\clearpage
\bibliographystyle{apalike} 
\bibliography{mybibilo.bib}